\documentclass[twocolumn,showpacs,preprintnumbers,amsmath,amssymb]{revtex4}

\usepackage[latin1, ansinew]{inputenc}       
\usepackage{graphicx}
\usepackage{dcolumn}
\usepackage{bm}

\begin{document}


\title{Rheophysics of dense granular materials :\\
Discrete simulation of plane shear flows}

\author{Frédéric da Cruz, Sacha Emam, Michaël Prochnow, Jean-Noël Roux and François Chevoir}
\email{Corresponding~author : chevoir@lcpc.fr} \affiliation{LMSGC,
Institut Navier, 2 allée Kepler, 77 420 Champs sur Marne, France}

\date{\today}

\begin{abstract}

We study the steady plane shear flow of a dense assembly of
frictional, inelastic disks using discrete simulation and
prescribing the pressure and the shear rate. We show that, in the
limit of rigid grains, the shear state is determined by a single
dimensionless number, called inertial number $I$, which describes
the ratio of inertial to pressure forces. Small values of $I$
correspond to the quasi-static regime of soil mechanics, while
large values of $I$ correspond to the collisional regime of the
kinetic theory. Those shear states are homogeneous, and become
intermittent in the quasi-static regime. When $I$ increases in the
intermediate regime, we measure an approximately linear decrease
of the solid fraction from the maximum packing value, and an
approximately linear increase of the effective friction
coefficient from the static internal friction value. From those
dilatancy and friction laws, we deduce the constitutive law for
dense granular flows, with a plastic Coulomb term and a viscous
Bagnold term. We also show that the relative velocity fluctuations
follow a scaling law as a function of $I$. The mechanical
characteristics of the grains (restitution, friction and
elasticity) have a very small influence in this intermediate
regime. Then, we explain how the friction law is related to the
angular distribution of contact forces, and why the local
frictional forces have a small contribution to the macroscopic
friction. At the end, as an example of heterogeneous stress
distribution, we describe the shear localization when gravity is
added.

\end{abstract}

\pacs{45.70.Mg, 81.05.Rm, 83.10-y, 83.80.Fg}

\maketitle

\section{Introduction} \label{sec:intro}

Due to their importance in geophysics (propagation of avalanches,
migration of dunes, fault sliding) and in various industrial
processes (handling of powders, granulates in civil and chemical
engineering, food, pharmacy, tribology...), flows of granular
materials are the focus of a large number of research, at the
frontier between physics and mechanics~\cite{Hutter94, Coussot99,
Rajchenbach00, Hinrichsen04}. In order to predict propagation,
flow-rate, jamming..., one of the main objectives of these
``rheophysical" studies is to determine the rheological laws of
those materials, based on their physical origin at the scale of
the grains and of their interactions. One thus tries to express
the stress tensor (and especially the pressure $P$ and the shear
stress $S$, positively counted) as a function of the shear rate
$\dot \gamma$ and other variables such as solid fraction $\nu$.

Granular materials are extremely various depending on the geometry
of the grains (shape and size) and the nature of their
interactions. We shall restrict our attention in the following to
assemblies of cohesionless grains, slightly polydisperse, without
interstitial fluid. This corresponds to macroscopic grains
(diameter larger than hundred microns) in a fluid of low viscosity
like air. The rheology is then only dictated by transfer of
momentum and dissipation of energy taking place in direct contacts
between grains and with the walls.

Depending on the conditions, these materials reveal various
mechanical behavior, similar to elastoplastic solids in the
quasi-static regime, to dense gazes in the case of strong
agitation, or to viscoplastic fluids when a flow is provoked. This
paper is devoted to this intermediate regime, which is still not
well understood. We shall first briefly recall the essential
results for the two extreme regimes, quasi-static and collisional.

\subsection{Quasi-static regime} \label{sec:qsreg}

Dense, confined granular assemblies in extremely slow shear flow
($\dot \gamma =10^{-4}\mbox{s}^{-1}$ for sand) are usually
described as solids abiding by elastoplastic constitutive
laws~\cite{Jackson83, Nedderman92, Vermeer98, Tardos97}, or more
general, incremental ones~\cite{Gudehus84, Wu96}. A typical
experiment consists in imposing a very slowly growing deviator
(shear) strain to a sample prepared in mechanical equilibrium
under an isotropic pressure. The material response is then
independent of physical time, and involves volumetric strain,
\emph{i.e.}, dilatancy~: loose samples contract, whereas dense
ones expand. In dense samples, the shear deviator stress reaches a
maximum (around shear strains of order $5\%$) and then decreases
to a plateau. In a loose system, it increases monotonically to the
same plateau value under the same confining stress. Meanwhile, the
solid fraction approaches a limit which does not depend on its
initial value either. It is therefore generally accepted that
after a large enough shear strain (of order $10\%$), the material
has reached a certain attractor state, called the \emph{critical
state}~\cite{Schofield68,Wood90}, which does not depend on its
initial arrangement. If larger shear strains are imposed, the
critical state, with its internal structure adapted to shear flow
in the limit of small velocities, will remain unaltered. The
critical state is characterized by an internal friction angle
$\phi$, defined as $\tan\phi = S/P$ in a simple shear test, and by
a critical solid fraction $\nu_{c}$. In the case of dense samples,
the observation of the critical state is often precluded by the
onset of shear localization, whereby the corresponding
microstructure is limited to the material that is continuously
sheared within thin bands.

Discrete numerical simulations have provided some additional
information on the internal structure of granular materials as
they approach their critical state. Thus the critical state is
associated with specific values for coordination numbers and
distribution of contact orientations (''fabric")~\cite{Radjai04}.
Simulations also contributed to unravel the influence of contact
elasticity on the macroscopic behavior. In the approach to the
critical state, strains are essentially due to rearrangements, as
contact networks break and repair. Macroscopic strains, with
typical pressure levels and contact stiffnesses, as soon as they
reach the $10^{-4}$ range, originate in geometry changes and not
in contact deformability~\cite{Roux02, Combe03}. As a sample
approaches its critical state, the gradually increasing anisotropy
of the contact network allows it to support increasing deviator
stresses~\cite{Radjai04}. Contact elasticity is essentially
irrelevant in that regime, which can be modelled with rigid
grains~\cite{Roux98}. The internal friction angle $\phi$ and the
critical state solid fraction $\nu_c$ are found to depend only
negligibly on the confining pressure $P$. The slight dependence
reported in experiments on sands~\cite{Wood90} is likely due to
features of the contact law, such as plasticity or breakage, which
are usually not introduced in numerical simulations. In other
words, the continuously sheared material can be modelled with the
Coulomb criterion, $S= P \tan  \phi $. If the \emph{direction} of
the imposed strain rate changes (\emph{e.g.}, upon reversing the
flow), then the system will leave its critical state and gradually
approach another one, with different contact
orientations~\cite{Radjai04}.

Quasi-statically deformed granular systems are one limit case for
the steady plane shear flows investigated in the present paper.
Therefore, the results will be presented in the sequel in terms of
an effective friction coefficient $\mu^*=S/P$, which should come
close to $\tan  \phi$ in the limit of slow motion. Likewise, as
appropriate for quasi-static granular rheology, we chose to
control the lateral pressure rather than keeping the solid
fraction fixed. Such options are further commented in the
following.

\subsection{Collisional regime} \label{sec:collreg}

In the dilute limit and/or for strong agitation, the grains
interact through binary, instantaneous, uncorrelated collisions.
Then, the generalization of the kinetic theory of dense gazes to
slightly inelastic grains~\cite{Haff83, Jenkins83, Jenkins85b,
Campbell90, Goldhirsch99} allows a hydrodynamical description. The
stress components depend on the solid fraction and on the velocity
fluctuations $\delta v$ (which square defines the so-called
\emph{granular temperature}). In the two-dimensional geometry
which we shall study in the following, the stress components are
homogeneous to a force divided by a length. For an assembly of
disks of diameter $d$ and mass $m$ :

\begin{eqnarray}
\label{eqn:kintheo1} \left \{ \begin{array}{cccc}
P &=&  F_P(\nu)  m (\delta v/d)^2, \\
\\
S &=&  F_S(\nu) m (\delta v/d) \dot{\gamma}.
\end{array} \right.
\end{eqnarray}

\noindent Solving a flow problem requires an additional equation
of energy in which a dissipation rate $\Gamma$ associated to
inelastic collisions must be added to the usual terms :

\begin{eqnarray}
\label{eqn:kintheo2} \Gamma &=&  F_{\Gamma}(\nu) m (\delta v/d)^3.
\end{eqnarray}

\noindent The dimensionless functions $F_i(\nu)$ are completely
expressed as functions of the pair correlation function at contact
$g_0(\nu)$. In the dense limit ($0.2 \le \nu \le 0.67$), $F_i(\nu)
\simeq A_{i} F(\nu)$, with $F(\nu) = \nu^2 g_0(\nu)$. In two
dimension, $g_0(\nu) = (16-7\nu)/16(1-\nu)^2$~\cite{Luding01b}.
The pre-factors $A_i$ are well-known functions of the restitution
coefficient $e$~\cite{Jenkins85b}.

In the case of an homogeneous system, where the shear rate and the
velocity fluctuations are uniform, the equation of energy reduces
to a balance between the work of the shear stress and the
dissipation : $S \dot{\gamma} = \Gamma$. This leads to $\delta v =
\dot \gamma d \sqrt{A_S/A_{\Gamma}}$. Consequently, the stress
components are equal to :

\begin{eqnarray}
\label{eqn:kintheo3} \left \{ \begin{array}{cccc}
P &=& G_P(\nu) m \dot{\gamma}^2, \\
\\
S &=& G_S(\nu) m \dot{\gamma}^2,
\end{array} \right.
\end{eqnarray}

\noindent where $G_{P,S}(\nu) = B_{P,S} F(\nu)$ with $B_{P} = A_P
A_S/A_{\Gamma}$ and $B_S = \sqrt{A_S^3/A_{\Gamma}}$. We notice
that the effective friction coefficient is a constant ($\mu^* =
\sqrt{A_SA_{\Gamma}/A_P^2}$), and that the solid fraction is a
function of the dimensionless quantity $I = \dot\gamma \sqrt{m/P}$
($\nu = G_P^{-1}(1/I^2)$). This collisional description is
relevant in the dilute limit when the inertial effects dominate.

\subsection{Dense regime}

In the intermediate regime, the solid fraction is close to a
maximum solid fraction, so that one speaks of \emph{dense flows}.
Then, the grains interact both through enduring contacts and
through collisions. There exists a contact network more or less
percolating through the material, which is very fluctuating in
space and time~\cite{Radjai02}. These flows are beyond the
quasi-static regime, since the inertia of the grains (and so the
shear rate) certainly comes into play. On the other hand, the
assumption of binary, instantaneous, uncorrelated collisions of
the kinetic theory is clearly in trouble. Due to the very strong
correlations of motion and force, the theoretical description of
those dense flows is very difficult and is still a matter of
debate (see~\cite{Pouliquen02a} for a recent review).

Advances have come in the last decade from the combination of
discrete numerical simulations and experiments on model materials
in simple geometry, confined or free surface flows (annular shear
cell, vertical chute, inclined plane, heap-flow), and in various
mechanical configurations (gravity, velocity or force imposed at a
wall). A detailed review of these works can be found
in~\cite{Gdr04}.

Depending on the mechanical configurations, the flows are steady,
intermittent, or even jam. A localization of the shear, with a
width of a few grains, is also frequently observed near the walls
or near the free surface, with exponential velocity profiles
around. However, the heterogeneity of the stress distribution as
well as the presence of walls makes the analysis of the
constitutive law difficult.

\subsection{Organization of the article}

This is the reason why we have chosen to study this dense flow
regime in steady homogeneous shear state. We have studied the
simplest geometry, plane shear without gravity, in which the
stress distribution is homogeneous inside the shear layer.
Furthermore, we have prescribed both the shear rate and the
pressure. Using discrete numerical simulations (``numerical
experiments"), we have access to microscopic information, at the
level of the grains and of the contact network, hardly measurable
experimentally, and we are able to vary the parameters describing
the grains and the shear state.

Sec.~\ref{sec:system} is devoted to the description of the
simulated system. Then in Sec.~\ref{sec:dimana} we build a
dimensionless number $I$, which, in the limit of rigid grains,
describes the shear state and characterizes the progressive
transition between the quasi-static and the dynamical regimes,
called intermediate regime. We then show in Sec.~\ref{sec:homog}
that we obtain homogeneous states in term of structure (solid
fraction), kinematics (shear rate) and stress distribution. In
Sec.~\ref{sec:behavior}, we measure the evolution of two
macroscopic quantities (solid fraction and effective friction
coefficient) as a function of $I$ in the intermediate regime, from
which we deduce the constitutive law. We then study the
sensitivity of this constitutive law on the mechanical properties
of the grains (restitution, friction and elasticity). The next
sections are devoted to more detailed information, first on the
fluctuations of the grain motion (Sec.~\ref{sec:fluctuation}), and
then on the contact network (Sec.~\ref{sec:network}). Then, using
those microscopic information, we explain in
Sec.~\ref{sec:frictionlaw} how the friction law is related to the
angular distribution of contact forces, and why the local
frictional forces have a small contribution to the macroscopic
friction. In the last Sec.~\ref{sec:gravity}, we study how the
previous results are affected when the stress distribution becomes
heterogeneous, in presence of gravity. For a more detailed account
of the results, we refer to~\cite{Dacruz04a}.

\section{Simulated system} \label{sec:system}

\subsection{Plane shear}

\begin{figure}[!htb]
        \begin{center}
            \includegraphics*[width=8cm]{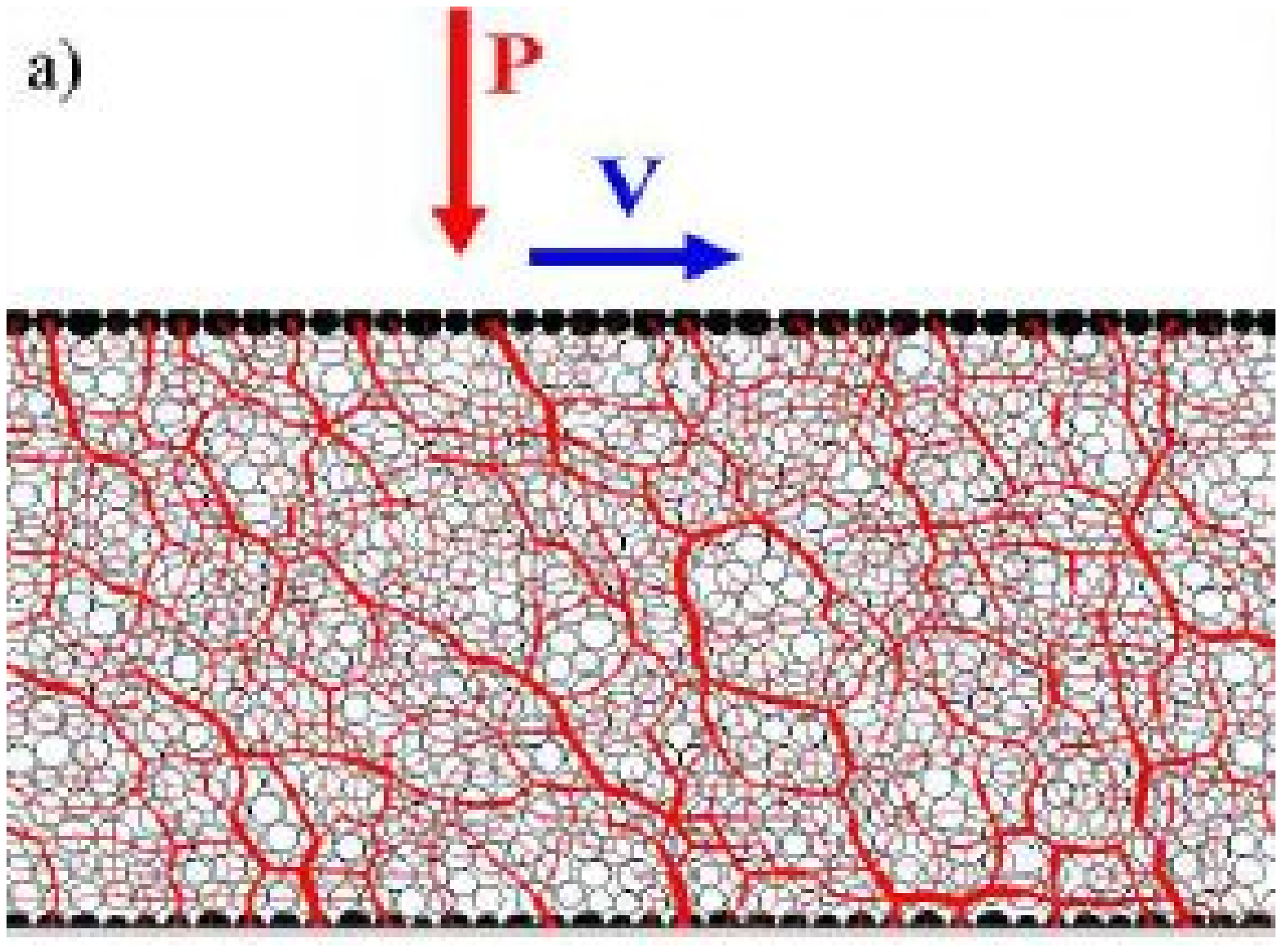}
            \includegraphics*[width=8cm]{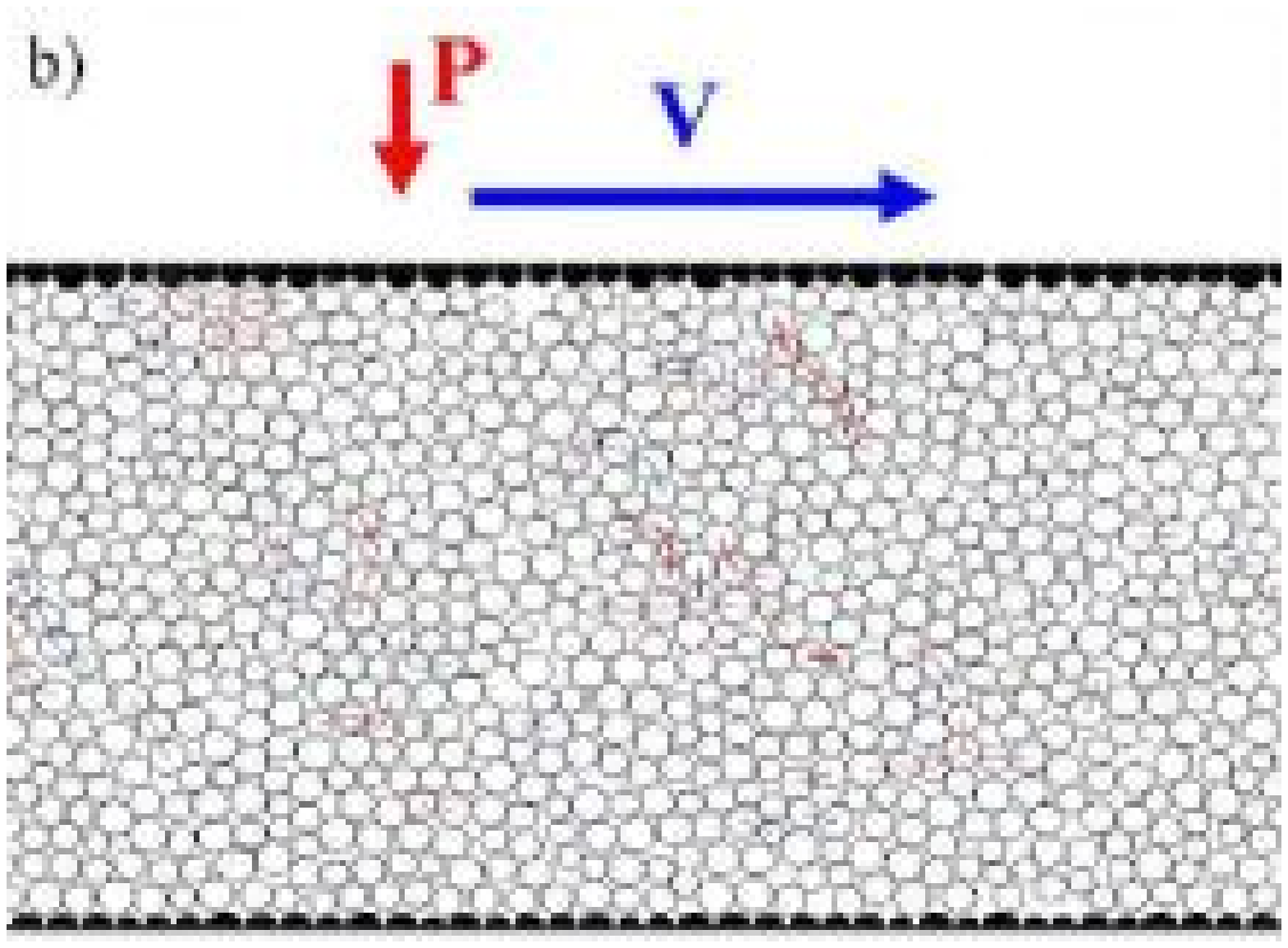}
            \caption{\label{Fig1}
            \textit{(Color online) Plane shear :
            (a) Quasi-static regime ($I=10^{-2}$) (b) Dynamic regime ($I=0.2$).
            (Black grains constitute the rough walls.
            The line widths are proportional to the intensity of the normal force between grains).}}
        \end{center}
    \end{figure}

The simulated system is two dimensional (Fig.~\ref{Fig1}). The
granular material is a dense assembly of $n$ dissipative disks of
average diameter $d$ and average mass $m$. Except in
Sec.~\ref{sec:rigid}, a small polydispersity of $\pm 20 \%$ is
considered to prevent crystallization~\cite{Luding01b}. The
mechanical properties of those grains are described by a friction
coefficient $\mu$, a restitution coefficient in binary collisions
$e$ and elastic stiffness coefficients $k_n$ and $k_t$.

The granular material is submitted to a plane shear, without
gravity (except in Sec.~\ref{sec:gravity}), so that the stress
distribution is uniform. The material is sheared between two
parallel rough walls, distant of $H$. One of the wall is fixed,
while the other moves at the prescribed velocity $V$. We call $x$
the flow direction and $y$ the transverse direction. Periodic
boundary conditions are applied along the flow direction, and we
call $L$ the length of the simulation box (always larger than $40$
grains). The wall roughness is made of contiguous grains sharing
the characteristics of the flowing grains (same polydispersity and
mechanical properties - no rotation). $y=0$ corresponds to the
center of the glued grains on the fixed wall.

An important feature of our simulation is that the normal stress
$\Sigma_{yy}$ is prescribed, as usual in practice. We shall see in
the following (Sec.~\ref{sec:stress}) that the normal stress
components $\Sigma_{xx}$ and $\Sigma_{yy}$ are equal, so that
$\Sigma_{yy}$ is equal to the pressure $P = (\Sigma_{xx} +
\Sigma_{yy})/2$. The control of the pressure $P$ is achieved by
allowing the dilatancy of the shear cell along $y$ ($H$ is not
fixed), through the vertical motion $\dot H$ of the moving wall.
The evolution of $H$ is given by : $\dot H = (P - P_w)L/g_{p}$,
where $g_p$ is a viscous damping parameter, and $P_w$ is the
normal stress exerted by the grains on the moving wall. Steady
state corresponds to $\langle P_w \rangle = P$.

The parameters of the simulated system are summarized in Tab.~ I
(anticipating on Sec.~\ref{sec:dimana}, we indicate the value of
the dimensionless parameters $g_{p}/\sqrt{mk_n}$ and $I$).

\begin{table}[!htb]
    \begin{center}
    \label{tab1}
    \begin{tabular}{c c c c c c}
        \\
        \hline
          & $n$ & $L/d$ & $H/d$ & $g_{p}/\sqrt{mk_n}$ & $I$\\
        \hline
        & 900 - 5000 & 40 - 100 & 20 - 100
        & 1 & $6. 10^{-4} - 0.3$\\
        \hline
    \end{tabular}
    \caption{\textit{List of system parameters.}}
    \end{center}
\end{table}

Due to its interest in rheology and more specifically in tribology
(third body) and in geophysics (sliding of faults at the origin of
earthquakes), this plane shear geometry has already been the
subject of numerous discrete simulation studies~\cite{Cundall89,
Babic90, Thompson91, Zhang92, Dent93, Lun94, Schwarz98,
Aharonov99, Morgan99, Hayakawa00, Jalali02, Aharonov02,
Campbell02, Radjai02, Volfson03, Iordanoff04, Lois05}, to which we
shall refer in the following. Those studies are generally
two-dimensional (except~\cite{Lun94, Campbell02, Jalali02}), with
a prescribed velocity (except ~\cite{Volfson03} where the shear
stress is prescribed), with a fixed volume
(except~\cite{Thompson91, Dent93, Aharonov02, Radjai02, Volfson03}
where the pressure is prescribed), and between rough walls (except
~\cite{Babic90, Lun94, Campbell02, Radjai02, Lois05} where
periodic boundary conditions are applied along the transverse
direction).

\subsection{Contact law} \label{sec:contactlaw}

Let us consider two grains $i$ and $j$ in contact, of diameter
$d_{i,j}$, mass $m_{i,j}$, centered at position ${\vec r}_{i,j}$,
with velocity ${\vec v}_{i,j}$ and rotation rate $\omega_{i,j}$.
Let ${\vec n}_{ij}$ denote the normal unit vector, pointing from
$i$ to $j$ (${\vec n}_{ij} = {\vec r}_{ij}/\vert \vert {\vec
r}_{ij}\vert \vert$ with the notation ${\vec r}_{ij} = {\vec r}_j
- {\vec r}_i$), and ${\vec t}_{ij}$ a unit tangential vector such
that $({\vec n}_{ij},{\vec t}_{ij})$ is positively oriented. We
call ${\vec F}_{ij}= N_{ij}{\vec n}_{ij}+T_{ij}{\vec t}_{ij}$ the
contact force exerted on the grain $j$ by the grain $i$. The
\emph{contact law} relates the normal, $N_{ij}$, and tangential,
$T_{ij}$, components of the contact force to the corresponding
components of relative displacements and/or velocities. The
relative velocity at the contact point is equal to ${\vec V}_{ij}
= {\vec v}_i -{\vec v}_j +1/2 (d_i\omega_i+d_j\omega_j) {\vec
t}_{ij}$. Its normal component $V_{ij}^N = {\vec n}_{ij}\cdot
{\vec V}_{ij}$ is the time derivative of the normal deflection of
the contact (or apparent ``interpenetration'' of undeformed
disks), $h_{ij} = (d_i + d_j)/2 - \vert \vert {\vec r}_{ij}\vert
\vert$. Its tangential component $V_{ij}^T={\vec t}_{ij}\cdot
{\vec V}_{ij}$ is the time derivative of the tangential relative
displacement $\delta _{ij}$. (Let us note that the definition of
$\delta _{ij}$ as a scalar quantity automatically accounts for the
material transport of strain in the contact region, as ${\vec
t}_{ij}$ moves with the pair in contact).

The normal contact force is the sum of two contributions, an
elastic one $N^e$ and a viscous one $N^v$ :
$N_{ij}=N^e_{ij}+N^v_{ij}$.

\subsubsection{Viscoelasticity}

Keeping in mind that contacts have to close to transmit forces
(${\vec F}_{ij}={\vec 0}$ if $h_{ij}<0$) the linear (unilateral)
elastic law reads

\begin{equation}
N^e_{ij} = k_n h_{ij}. \label{eqn:Ne}
\end{equation}

\noindent Eqn.~\eqref{eqn:Ne} involves a constant normal stiffness
coefficient $k_n$, the value of which is independent of disk
radii.

The normal viscous force opposes the relative approaching or
receding velocity~:

\begin{equation}
N^v_{ij} = \zeta_{ij} \dot h_{ij}. \label{eqn:Nv}
\end{equation}

\noindent $\zeta_{ij}$ is related to the normal restitution
coefficient $e$ in a binary collision, and chosen such that $e$ is
constant for all contacting pairs, whence

\begin{equation}
\zeta _{ij} = \frac{-2\ln e}{\sqrt{\pi^2+\ln^2
e}}\sqrt{\frac{m_{ij}}{k_n}}. \label{eqn:restit}
\end{equation}

\noindent where $m_{ij} = m_i m_j/(m_i+m_j)$.

Physically, Eqn.~\eqref{eqn:Ne} can be regarded as a simplified
version of the Hertz law~\cite{Johnson85}, $N^e\propto h^{3/2}$,
while the viscous dissipation might stem from the visco-elasticity
of the grain material~\cite{Brilliantov96b}.

The total normal force might be either repulsive or attractive,
due to the viscous contribution. We could check that setting
$N_{ij}$ to zero whenever it becomes attractive ($N_{ij}<0$) has
but a negligible effect on the simulation
results~\cite{Schollmann99}.

\subsubsection{Friction}

The Coulomb condition in the contacts involves the coefficient of
friction between grains $\mu$, and is enforced with the sole
elastic part of the normal force~:

\begin{equation}
\vert T_{ij}\vert \le \mu N_{ij}^e. \label{eqn:T1}
\end{equation}

\noindent To this end, the tangential component of the contact
force is related to the \emph{elastic part} $\delta _{ij}^e$ of
the relative tangential displacement $\delta _{ij}$,

\begin{equation}
T_{ij} = k_t \delta _{ij}^e, \label{eqn:T2}
\end{equation}

\noindent with a tangential stiffness coefficient $k_t$. $\delta
_{ij}^e$ is defined by~:

\begin{equation}
\frac{d\delta _{ij}^e}{dt}=\begin{cases} 0& \text{if $\vert
T_{ij}\vert = \mu N_{ij}^e$ and $T_{ij}
  V_{ij}^T>0$}\\
V_{ij}^T & \text{otherwise}
\end{cases}
\label{eqn:T3}
\end{equation}

The contact is termed ``sliding'' in the first case in
Eqn.~\eqref{eqn:T3} (the condition that $T_{ij}$ and $ V_{ij}^T$
share the same sign ensuring a positive dissipation due to
friction) and ``rolling'' in the second case. $k_t$ is of the same
order of magnitude as $k_n$~\cite{Johnson85}. As it has a very
small influence on the results~\cite{Campbell02}, it was fixed to
$k_n/2$ in all our calculations.

Table~\ref{tab:material} gives the list of material parameters
(anticipating on Sec.~\ref{sec:dimana}, we indicate the value of
the dimensionless parameter $\kappa$). Most simulations were done
with $e = 0.1$ and $\mu = 0.4$, which implies a rather strongly
dissipative material, favors dense flows, and seems fairly
realistic. Other friction coefficients and restitution
coefficients were also studied.

\begin{table}
    \begin{tabular}{c c c c c c}
        \\
        \hline
          & polydispersity & $\mu $ & $e$ & $k_t/k_n$ & $\kappa$ \\
        \hline
        & $\pm 20 \%$ & $0$--$0.8$ & $0.1$--$0.9$ & $0.5$ & $10^4$\\
        \hline
    \end{tabular}
    \caption{\textit{List of material parameters.}}
    \label{tab:material}
\end{table}

\subsection{Simulation method}

The interaction law being chosen, numerical simulations are
carried out with the molecular dynamics method, as in
refs.~\cite{Cundall79, Silbert01, Roux05}. The equations of motion
are discretized using a standard procedure (Gear's order three
predictor-corrector algorithm~\cite{Allen87}). The time step is a
small fraction ($1/100$) of the duration $\tau_{c}$ of a binary
collision between two grains of mass $m$ ($\tau_{c} =
\sqrt{m(\pi^{2} +\ln^2 e)/(4 k_{n})}$).

    \section{Dimensional analysis} \label{sec:dimana}

In order to analyze the results, we recall the list of parameters
describing the material and the shear state. The grains are
described by their size $d$, mass $m$, stiffness parameters $k_n$
and $k_t$, and coefficients of restitution $e$ and of friction
$\mu$. The shear state is described by the prescribed pressure $P$
and the average shear rate $V/H$, and by the viscous damping
parameter $g_p$. In the following, we shall express the results in
the natural units of the simulated system. If the length and mass
scales $d$ and $m$ are obvious, there are three candidates for the
time scale : the shear time $1/\dot \gamma$, the inertial time
$\sqrt{m/P}$ (that is to say the characteristic displacement time
of a grain of mass $m$ submitted to a pressure $P$), and the
collision time $\tau_c$. However, dimensional analysis predicts
that the behavior will only depend on six dimensionless numbers.
Apart from $\mu$, $e$ and $k_t/k_n$, we propose the following
choice of the three other dimensionless numbers:

\begin{enumerate}

    \item[i)] The first $g_{p}/\sqrt{mk_{n}}$ is associated to the normal
    motion of the wall controlling the pressure. A small value signifies
    that the time scale of the fluctuations of $H$ is imposed by the material
    rather than the wall, and that the wall ``glues" to the material.

    \item[ii)] The second $\kappa = k_n/P$ is associated to the rigidity
    of the grains. It is inversely proportional to the normal deflection $h$
    of the contacts for a confining pressure $P$.

    \item[iii)] The third $(V/H)\sqrt{m/P}$ describes the shear state, through
    a combination of the three global parameters (velocity $V$, pressure $P$ and height
    $H$), which means that it is not necessary to vary independently those three
    parameters.

\end{enumerate}

The first dimensionless number is equal to $1$ in our simulations.
We have not studied its influence, but consider that is is small
in this range. We shall not refer to it in the following.

Most of our calculations are restricted here to the limit of rigid
grains~\cite{Lois05}, $\kappa\to \infty$, as we chose $\kappa =
10^4$ (unless specified otherwise, as in Sec.~\ref{sec:rigid}).
From studies in the quasi-static regime, this value is known to be
large enough for the coordination number to show little variation
as $\kappa$ is further increased (see, however,
Sec.~\ref{sec:coordination}). With $\kappa = 10^4$, the typical
ratio $h/d \propto \kappa^{-1}$ is so small that elastic
deflections $h$ stay negligible in comparison with the gaps
between neighboring grain surfaces that determine the amplitude of
rearrangement events~\cite{Roux02} (a more stringent condition
than $h\ll d$). In practice, a suitable definition of $\kappa$ for
particles with Hertzian contacts (such that $h/d \propto
\kappa^{-1}$ with a coefficient of order 1) is $\kappa = (E/P)
^{2/3}$, where $E$ is the Young modulus of the material the grains
are made of. For glass ($E = 70$ GPa) and $P = 10kPa$ (the
pressure due to the weight of a $50$ cm thick layer) one has
$\kappa\simeq 37000$.

\subsection{Inertial number $I$}

In the following, we study in detail the influence of the third
dimensionless number. It is a global number $I_g$, at the scale
$H$ of the sheared layer. Replacing the global shear rate $V/H$ by
the local one $\dot{\gamma}$ defines the local analogous quantity
:

\begin{equation} \label{eqn:defI}
    I = \dot{\gamma} \sqrt{\frac{m}{P}}.
\end{equation}

As the ratio of inertial to shear times, $I$ measures the inertial
effects, and will be called \emph{inertial number} in the
following. This number already appeared in the collisional regime
(Sec.~\ref{sec:collreg}). In a three dimensional situation, the
inertial number is equal to $\dot{\gamma}\sqrt{m/Pd}$. We notice
that, introducing the mass density $\rho_g$ of the grains, the
definition $I = \dot{\gamma} d \sqrt{\rho_g/P}$ does not depend on
the dimensionality of the system.

In a homogeneous system, without sliding velocity at the wall,
both definitions, global and local, are equivalent. But in a
heterogeneous system, it is necessary to take into account the
variations of the local inertial number along $y$: $I(y) =
\dot{\gamma}(y) \sqrt{m/P(y)}$.

We are going to show that this dimensionless number might well be
the fundamental quantity to describe the rheology of granular
materials. First of all, it is the correct quantity to
characterize the flow regime. Thus, the terminology of ``slow"
versus ``rapid" granular flows is not correct, since it is
necessary to combine shear rate and pressure to characterize the
inertial effects. A small value of $I$ (small $\dot\gamma$ and/or
large $P$) corresponds to a regime where the grain inertia is not
relevant : this is the ``quasi-static" regime (Fig.~\ref{Fig1}
(a)). Inversely, a large value of $I$ (small $P$ and/or large
$\dot\gamma$) corresponds to the ``inertial" or ``dynamical"
regime, which may be described by the kinetic theory
(Fig.~\ref{Fig1} (b)). Varying $I$ allows to study the progressive
transition between those two regimes. The range of inertial number
which we have studied goes from $6 \cdot 10^{-4}$ to $0.3$.

\subsection{Comments} \label{sec:dimanacom}

It might be pointed out that the inertial number is directly
related to the \emph{Bagnold number} $Ba$, which can be defined as
the ratio of the typical kinetic energy $m(d\dot\gamma)^2$ of one
grain to the characteristic frictional dissipation $\mu
Pd^2$~\cite{Coussot99}. Some authors~\cite{Babic90, Campbell02}
chose to control the solid fraction $\nu$ rather than the
pressure, and therefore, rather than $I$ and $\kappa$, used the
pair of dimensionless numbers $\nu$ and
$\alpha=\dot{\gamma}/\sqrt{k_n/m} = I/\sqrt{\kappa}$, as variables
characterizing the state of the granular material in steady
homogeneous shear flow. The latter dimensionless combination
$\alpha$ may be viewed as the ratio of the collision time to the
shearing time, or as the shearing velocity divided by the sound
velocity (Mach number~\cite{Campbell02}). Both choices are
perfectly legitimate, as dimensional analysis predicts either $S/P
= f_1(I,\kappa)$ and $\nu = f_2(I,\kappa)$, or $S/P =
f_3(\nu,\alpha)$ and $P/k_n = f_4(\nu,\alpha)$, both results being
equally valid. The choice of $I$ and $\kappa$ can however be
deemed more convenient for several reasons. First, the variation
of the results with $\nu$, regarded as a control parameter, is
extremely fast. Each material possesses a critical packing
fraction $\nu_c$, in the sense of Sec.~\ref{sec:qsreg}, above
which it does not flow, unless stresses are so large that the
elastic compression of contacts compensates for the difference
$\nu-\nu_c$. Below $\nu_c$, on the other hand, a continuously
sheared granular system is free to flow with a negligible shear
stress, unless the velocity is high enough to build a significant
pressure. One should therefore monitor $\nu$ with great accuracy
to observe ordinary stress levels. This renders the comparisons
between different granular systems difficult, as one would need to
know in advance the value of the critical density for each of
them. Furthermore, the limit of rigid grains becomes singular, as
all values of $\nu$ above $\nu_c$ are strictly forbidden for $\dot
\gamma \neq 0$, while the properties of shear flows with
$\nu<\nu_c$ simply scale with $\dot \gamma$ in that
limit~\cite{Lois05}. Conversely, if one uses $I$ and $\kappa$ as
control parameters, no singularity enters any of the relevant
results in the $I\to 0$ or $\kappa\to \infty$ limits, and
different materials should exhibit similar (if not quantitatively
identical) behaviors for the same values of these parameters
(which thus define roughly ``corresponding states''). It should
also be pointed out that experimental conditions usually determine
stress levels, rather than densities.

\section{Homogeneous shear state}
\label{sec:homog}

\subsection{Preparation of steady shear states}

The first kind of preparation (which has been used most of the
time) consists in starting from an initial configuration where the
disks are randomly deposited without contact and at rest between
the two distant walls, which provides an average solid fraction of
$0.5$, and then in applying the pressure to the wall while slowly
shearing the granular material. When the pressure on the walls
reaches the prescribed value, the prescribed velocity is applied.

The second kind of preparation consists in starting from a very
high solid fraction (of the order of $0.8$), obtained by a random
deposit followed by a cyclic compaction with frictionless grains,
and then to introduce friction between grains and to start the
shear with the prescribed velocity and pressure.

The two kinds of preparation allow to start either from a loose
state (the first case), or from a dense state (the second case).
Occasionally, we have used a third kind of preparation consisting
in starting from localized shear states near one of the walls
(obtained by applying gravity, see Sec.~\ref{sec:gravity}).

We then look for a steady flow, characterized by constant
time-averaged quantities of the flowing layer, like kinetic energy
and solid fraction. We have observed that, after a sufficient
amount of time, the three kinds of preparation lead to the same
shear state. We deduce that there is no influence of the
preparation on the steady flow characteristics.

All the simulations converge to an average steady state. But the
relative fluctuations of the measured quantities can be very
different. In the following, we consider that the shear state is
continuous if they are smaller than $10\%$. Otherwise, the flow is
called intermittent. This happens in the quasi-static regime, for
$I \le 0.001$ (see Sec.~\ref{sec:intermit}).

When a steady state is reached, the simulation is carried on
during a sufficient amount of time $\Delta t$, so that the typical
relative displacement of two neighboring layers is larger than ten
grains ($\dot \gamma \Delta t \ge 10)$. In this steady state, we
consider that the statistical distribution of the quantities of
interest (structure, velocities, forces\dots) are independent of
$t$ and $x$, so that we average both in space (along $x$) an in
time (considering $200$ time steps distributed over the period
$\Delta t$).

\subsection{Profiles}

\begin{figure}[!htb]
        \begin{center}
            \includegraphics*[width=8cm]{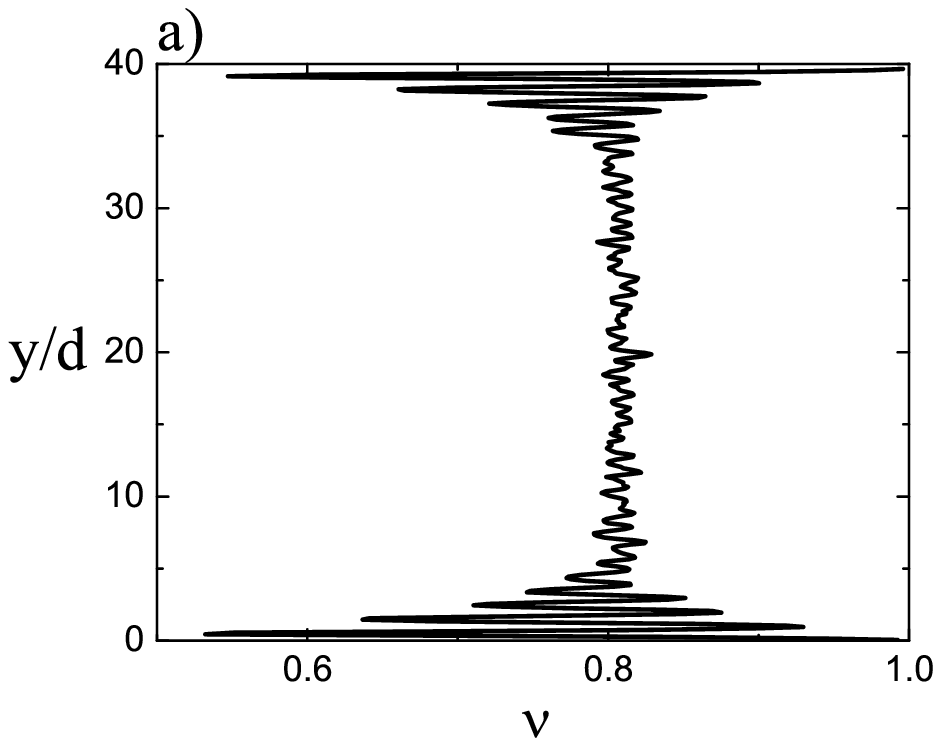}
            \includegraphics*[width=8cm]{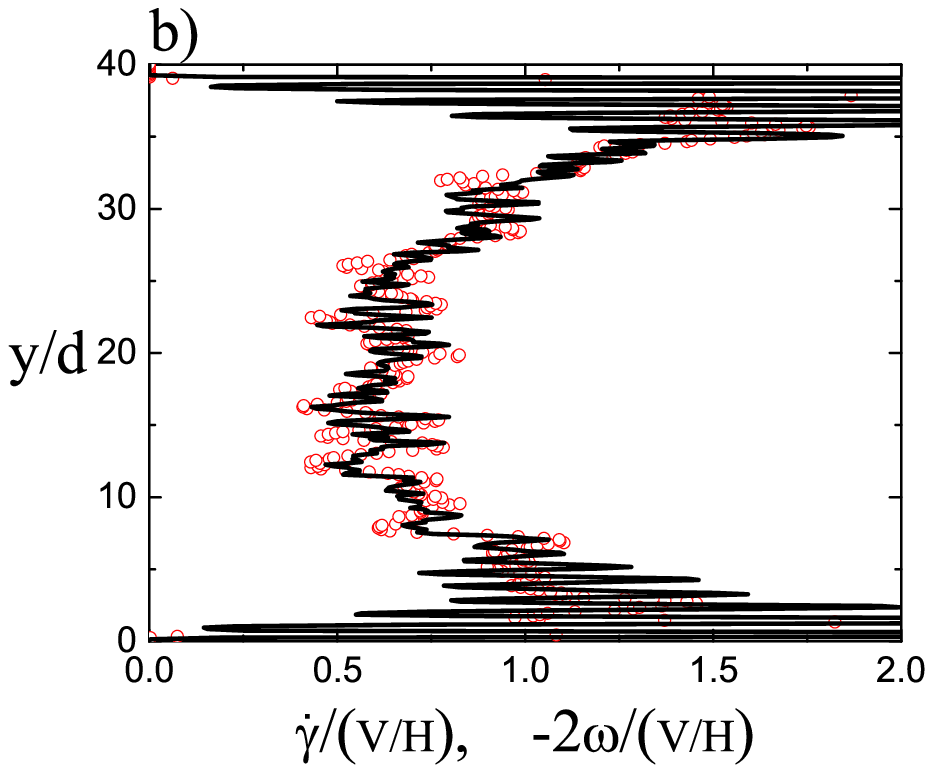}
            \includegraphics*[width=8cm]{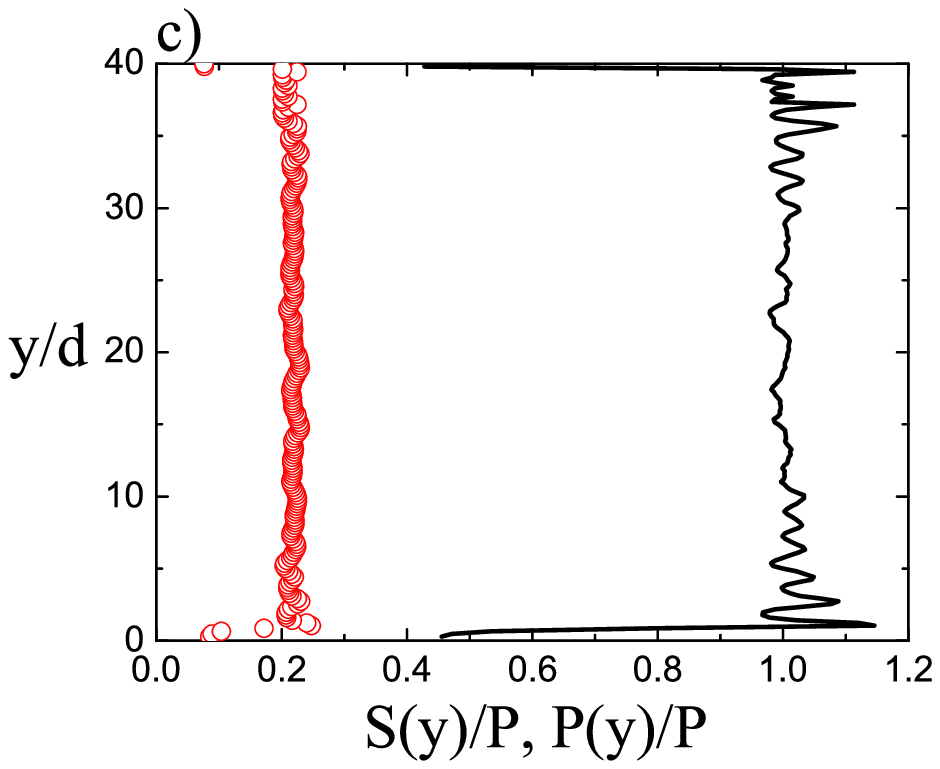}
            \caption{\label{Fig2}
            \textit{(Color online) Homogeneous shear state : (a) Solid fraction $\nu (y)$, (b) Shear rate $\dot \gamma (y)$ ($-$)
            and rotation velocity $-2\omega(y)$ ($\circ$), (c) Pressure $P(y)$
            ($-$) and shear stress $S(y) $ ($\circ$) ($V = 1$,$P = 25$, $H/d \approx 40$, $e = 0.1$, $\mu =
0.4$).}}
        \end{center}
    \end{figure}

We now show that the granular material is completely sheared and
that the shear is homogeneous. To this end, we analyze the
profiles of solid fraction, velocity and stress components.
Typical profiles are shown in Fig.~\ref{Fig2}.

\subsubsection{Solid fraction and velocity} \label{sec:nuvprofiles}

The profiles of solid fraction $\nu(y)$ and shear rate
$\dot{\gamma}(y)$ are strongly oscillating near the walls, but we
may consider them as approximately constant in the central part of
the sheared layer (Fig.~\ref{Fig2} (a) and (b)). This allows to
define average solid fraction $\nu$ and shear rate $\dot \gamma$.

We notice that those profiles obtained for controlled pressure are
in agreement with the one measured at fixed
volume~\cite{Dacruz04b}.

We have also measured the profile of average rotation rate $\omega
(y)$ (Fig.~\ref{Fig2} (b)), and observed, as in other flows and
quasi-static deformations ~\cite{Lun94, Calvetti97, Azanza98,
Howell99b, Latzel00, Prochnow02}, the following relation
$\omega(y) = - \frac{1}{2} \dot{\gamma}(y)$.

\subsubsection{Stress tensor} \label{sec:stress}

The stress tensor $\underline{\underline{\Sigma}}$ is the sum of
two contributions~\cite{Batchelor70}:

\begin{equation}
    \underline{\underline{\Sigma}} = \underline{\underline{\Sigma}}^{c} +
    \underline{\underline{\Sigma}}^{f}.
\end{equation}

The first term (``contact"), usual in static of granular
materials, is associated to contact forces between
grains~\cite{Christoffersen81, Kruyt96}. The second term
(``fluctuations"), usual in fluid mechanics (Reynolds tensor), is
associated to the velocity fluctuations of the grains $\delta \vec
v_{i}$~:

\begin{equation}
\left \{ \begin{array}{cccc}
    \underline{\underline{\Sigma}}^{c} & = & \frac{1}{LH}
    \sum_{i<j} \vec F^{ij} \otimes \vec
    r^{ij}\\
    \\
    \underline{\underline{\Sigma}}^{f} & = & \frac{1}{LH}
    \sum_{i=1}^{n} m_{i} \delta \vec v_{i} \otimes \delta \vec
    v_{i}
\end{array} \right.
\label{eqn:sigma}
\end{equation}

A third contribution, associated to the rotation of the grains,
has been introduced in~\cite{Moreau97}. We have observed that is
is usually insignificant~\cite{Dacruz04a}, so that we shall not
discuss it in the following. We discuss here the total
contribution and shall analyze the two contributions in
Sec.~\ref{sec:fluctuation}.

Using the same averaging methods than for solid fraction and
velocities~\cite{Prochnow02}, it is possible to calculate the
profiles of the different components of the stress tensor. We
observe that $\Sigma_{xx} \simeq \Sigma_{yy}$, like for dense
flows down inclined planes~\cite{Prochnow02, Silbert01, Lois05}.
Consequently the pressure $P = (\Sigma_{xx}+\Sigma_{yy})/2 \simeq
\Sigma_{yy}$. In the following we call $S = - \Sigma_{xy}$. The
Fig.~\ref{Fig2} (c) shows the profiles $P(y)$ and $S(y)$. Those
components are approximately constant along $y$, as expected.

\subsubsection{Inertial number}

From the profiles of shear rate and pressure inside the flow, we
deduce the profile of inertial number $I(y)$. It is approximately
constant in the center of the sheared layer, so that it is
possible to define an average $I$ (in Fig.~\ref{Fig2} $I \approx
0.004$).

\subsection{Behavior near the walls}\label{sec:walls}

Particular behaviors of solid fraction, velocity and stress
components are observed near the walls. First, the walls induce a
structuration of the granular material in the $5$ first layers,
evidenced by the oscillations of the solid fraction and shear
rate. Second, the frustration of the rotation of the flowing
grains in contact with the rotationless glued grains of the
roughness induces a sudden variation of the average rotation. This
may be responsible for the difference of normal stress components
($\Sigma_{xx} \neq \Sigma_{yy}$) in the first two layers (not
shown). Third the granular material is agitated by the roughness,
which increases the shear rate (see Sec.~\ref{sec:fluctuation}).
Consequently, all the average quantities are measured in the
central part of the sheared layer, excluding the $5$ first layers
near the walls and $H$ is chosen large enough so as to limit those
wall effects.

However, we have tested the homogeneity of the shear in the case
of a thin layer ($H/d \approx 5$)~\cite{Dacruz04a}. Then the
granular material is structured on its whole width and completely
sheared. We shall come back to this case in the following.

\subsection{Limits of homogeneity} \label{sec:limit}

Even when starting from a localized velocity profile, we have
observed a relaxation toward a homogeneous shear state. This
observation is in contrast with other studies, where a shear
localization is observed~\cite{Cundall89, Morgan99, Jalali02,
Aharonov02}. In our simulations, we have observed signs of
localization in the quasi-static regime where the flow becomes
intermittent (Sec.~\ref{sec:fluctuation}), and once the stress
distribution becomes heterogeneous (Sec.~\ref{sec:gravity}). We
now briefly discuss two other factors of localization, which are
associated to the granular material itself, the monodispersity and
the softness of the grains.

\subsubsection{Influence of the polydispersity} \label{sec:polydispersity}

In the case of a very small polydispersity ($\le 1\%$), we have
observed~\cite{Dacruz04a} that the granular material crystallizes
near the walls. The shear zone then reduces to a ten diameters
thick central layer, where the velocity profile is linear. When
the polydispersity is increased, the shear zone extends first to
the moving wall ($5\%$) then to the whole layer ($10\%$).

\subsubsection{Influence of the rigidity} \label{sec:rigid}

We expect that the correlation length of the strain field becomes
smaller when the grains become softer. This might be responsible
for a localization of the shear near the moving wall. In order to
test this idea, we have studied the influence of the dimensionless
rigidity number $\kappa$ in a large system ($H/d = 100$) for
$I_{g} = 2 \cdot 10^{-4}$. For $\kappa = 1000$, we observe
homogeneous shear states, whereas for $\kappa = 40$, the shear
becomes localized~\cite{Dacruz04a}.

\section{Constitutive law} \label{sec:behavior}

Those homogeneous states are a great advantage to measure the
constitutive law of dense granular flows. This would be much more
difficult in other flow geometries (annular shear, vertical chute,
inclined plane, heap-flow\dots)~\cite{Gdr04} where the flow is
heterogeneous, so that the constitutive law cannot be identified
on the whole flowing layer, but should be studied locally, in the
presence of gradients.

In those homogeneous states where the inertial number $I$ is
prescribed, the solid fraction and the shear stress adjust in
response to $I$. Consequently, we shall first measure the
dependences of two fundamental dimensionless quantities, the solid
fraction and the effective friction coefficient, as functions of
$I$. Then we will study the influence of the mechanical properties
of the grains, the coefficients of restitution $e$ and friction
$\mu$.

    \subsection{Dilatancy law}

\begin{figure}[!htb]
        \begin{center}
            \includegraphics*[width=8cm]{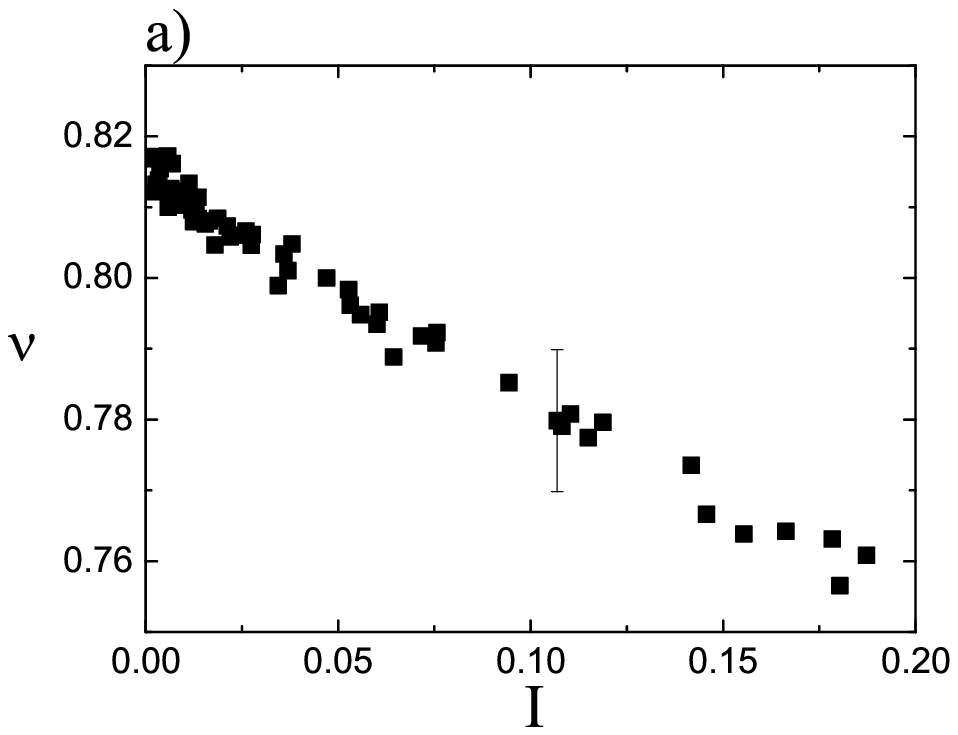}
            \includegraphics*[width=8cm]
            {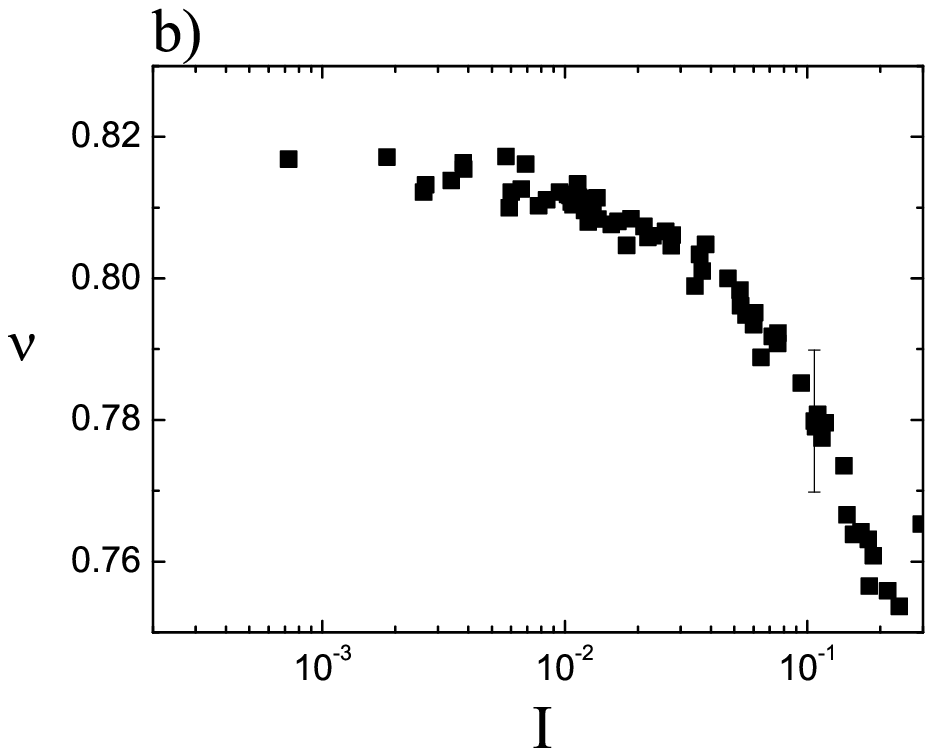}
            \caption{\label{Fig3}
            \textit{Dilatancy law :
            (a) Linear scale (b) Logarithmic scale
            ($\mu = 0.4$, various $e$ and $\kappa$).}}
        \end{center}
    \end{figure}

We call ``dilatancy law" the variations of the average solid
fraction $\nu$ as a function of the inertial number $I$
(Fig.~\ref{Fig3}). We observe that $\nu$ decreases approximately
linearly with $I$, starting from a maximum value $\nu_{max}$ :

\begin{equation}
    \label{eqn:nu(i)}
    \nu(I) \simeq \nu_{max} - aI,
\end{equation}

\noindent with $\nu_{max} \simeq 0.81$ and $a \simeq 0.3$ (for
$\mu = 0.4$). The error bar (independent of $I$) corresponds to
the statistical dispersion inside the layer.

    \subsection{Friction law}

The effective friction coefficient has been defined as the ratio
of the shear stress to the pressure inside the material $\mu^{*} =
S/P$.

It could also be defined as the ratio of the (total) tangential
and normal forces on the wall $\mu_{w}^{*} = T/N$. We have
observed~\cite{Dacruz04a} that $\mu_{w}^{*}$ is slightly larger
than $\mu^{*}$. Some simulations have been carried out on to test
the influence of the roughness, by taking glued grain on the wall
twice as small ($R = 0.5$) or twice as large ($R = 2$) as the
flowing grains, for the same $I_g$. This size ratio has an
influence on the sliding velocity at the wall : it becomes
noticeable for $R = 0.5$ and decreases when $R$ increases, since
the grains close to the walls are trapped by the roughness.
However, at distance from the walls, the flow remains homogeneous,
but the shear rate, and hence $I$, decreases when $R$ decreases.
Furthermore, the effective friction at the wall decreases when $R$
decreases. All in all, $\mu_{w}^{*}(I)$ seems independent of $R$.
For a more detailed discussion of the influence of the roughness
on the flow (inclined plane and vertical chute), we refer
to~\cite{Prochnow02, Goujon03, Gdr04}. In the following we shall
only discuss the effective friction coefficient in the volume of
the flowing layer.

\begin{figure}[!htb]
        \begin{center}
            \includegraphics*[width=8cm]{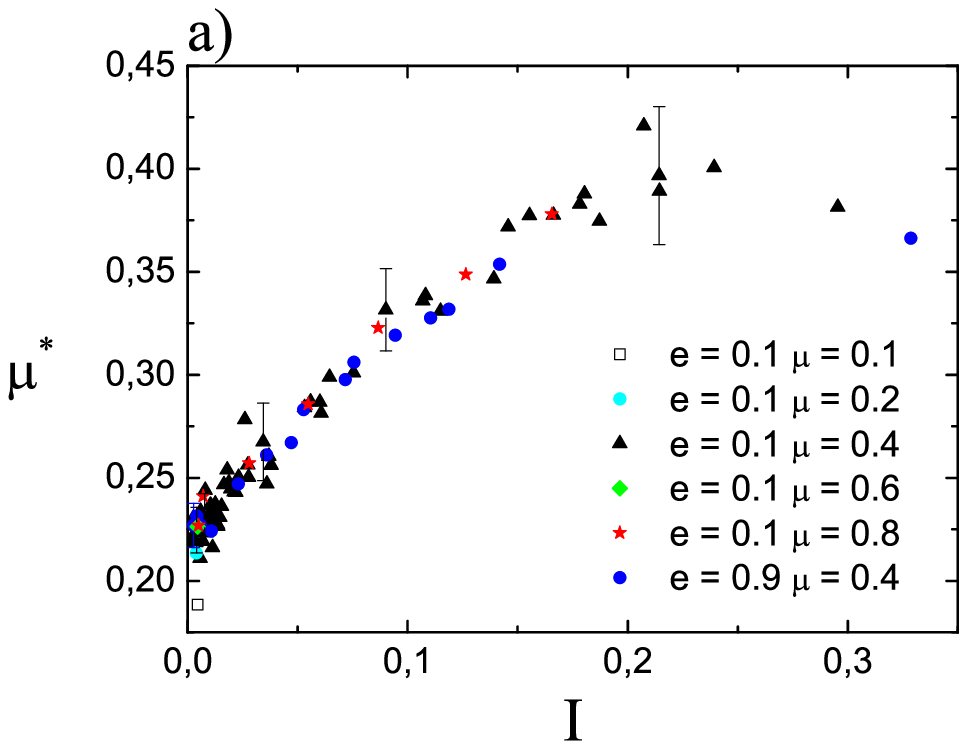}
            \includegraphics*[width=8cm]{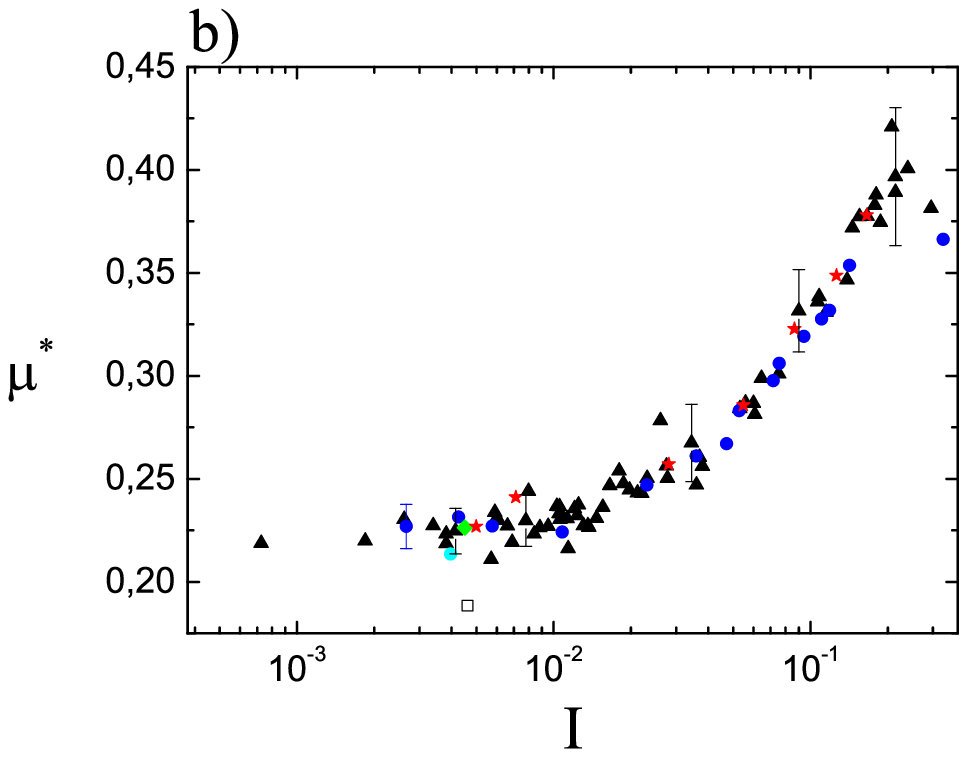}
            \caption{\label{Fig4}
            \textit{(Color online) Friction law : (a) Linear scale (b) Logarithmic
            scale ($\mu \neq 0$, various $e$ and $\kappa$).}}
        \end{center}
    \end{figure}

We call ``friction law" the variations of the effective friction
coefficient $\mu^{*}$ (averaged over the width in the central part
of the flowing layer) as a function of $I$ (Fig.~\ref{Fig4}). We
observe that $\mu^{*}$ increases approximately linearly with $I$,
starting from a minimum value $\mu^{*}_{min}$ :

\begin{equation}
    \label{eqn:mu(i)}
    \mu^{*} \simeq \mu^{*}_{min} + bI,
\end{equation}

\noindent with $\mu^{*}_{min} \simeq 0.22$ and $b \simeq 1.0$ (for
$\mu \neq 0$). The error bars (independent of $I$) correspond to
the statistical dispersion inside the layer. We also observe that
$\mu^{*}$ tends to saturate for $I \ge 0.2$. Within the error
bars, it is difficult to be more precise about those dependencies.
A more careful measurement is deferred for future work.

We now compare this friction law with other works. We first notice
that the increase of $\mu^{*}$ with $I$ is contrary to the well
documented decrease of the friction coefficient with the velocity
in the quasi-static regime~\cite{Chambon03}. However, in those
studies, this softening is interpreted as a consequence of the
renewal of the population of asperities at a microscopic scale, or
as an effect of humidity~\cite{Ovarlez03}. Those effects do not
come into play in our study. As a matter of fact, this friction
law was already observed in previous discrete simulations
~\cite{Dent93}, and partial observations (experimental or
numerical) of the variation of $\mu^{*}$ with the shear rate, the
pressure or the solid fraction, consistent with our observations,
may be found in~\cite{Savage84, Hanes85b, Craig86, Thompson91,
Campbell95, Potapov96, Morgan99, Klausner00, Ovarlez03}.
Interestingly, the inclined plane geometry allows to prescribe
both the effective friction and the pressure, through the
inclination $\theta$ of the plane and the height $H$ of the
flowing layer. Consequently, the measure of the superficial
velocity $V$ as a function of these two parameters provides a
measure of the friction coefficient at the base as a function of
$I_g$ (which is proportional to $V/H^{3/2}$) ~\cite{Pouliquen99a,
Pouliquen02b}. Those observations are in good agreement with the
previous friction law~\cite{Dacruz04a, Dacruz05}.

    \subsection{Comments}

As a conclusion, in the limit of rigid grains, the classification
of the flow regimes depends on the single dimensionless number
$I$. In the quasi-static regime ($I \le 10^{-2}$), the granular
material is very dense, close to the maximum solid fraction
$\nu_{max}$, and the effective friction coefficient is close to
its minimum value $\mu^*_{min}$. In the dynamic regime ($I \ge
0.2$), the dilatancy becomes strong and the effective friction
coefficient seems to saturate. The transition between those two
regimes is progressive. In the intermediate regime ($10^{-2} \le I
\le 0.2$), we observe approximately linear variations of the solid
fraction and of the effective friction coefficient as a function
of $I$ (Eqs.~\eqref{eqn:nu(i)} and~\eqref{eqn:mu(i)}).

In the case of a thin layer (see Sec.~\ref{sec:walls}), the
dilatancy and friction laws are not affected. The parameters
remain the same, except for the solid fraction which is smaller
than for a thick layer : $\nu_{max} = 0.82$ instead of $0.84$ for
$\mu=0$ and $\nu_{max} = 0.78$ instead of $0.81$ for $\mu=0.4$.

In the case of a very small polydispersity (see
Sec.~\ref{sec:limit}), the friction law is preserved in the
central sheared layer. For larger polydispersity ($\le 50\%$), the
dilatancy and friction laws are not affected.

When taking into account the elasticity of the grains, it is
natural to draw a diagram of the flow regimes as a function of the
two dimensionless numbers, already quoted in
Sec.~\ref{sec:dimanacom}, $\alpha = I/\sqrt{\kappa}$ and
$\nu$~\cite{Babic90, Campbell02}. This leads to identify three
regimes : elastic quasi-static, purely inertial and
elastic-inertial.  This last regime corresponds to very soft
grains ($\kappa < 100$) and is not accessible in our study, where
we stay in the limit of rigid grains.

Furthermore, various studies where the volume rather than the
pressure was prescribed~\cite{Babic90, Aharonov99, Howell99b,
Campbell02} have evidenced a transition between the quasi-static
and the inertial regimes around a critical solid fraction. In our
study where the pressure is prescribed, the solid fraction adjusts
to the inertial number $I$, so that the transition is not
accessible.

\subsection{Constitutive law}

We shall now show how these dilatancy and friction laws allow to
deduce the constitutive law of the material, that is to say the
dependencies of the pressure and shear stress on the shear rate
and solid fraction :  $P(\nu,\dot\gamma)$ and $S(\nu,\dot\gamma)$.

        \subsubsection{Pressure}

From the definition of $I$ (Eqn.~\eqref{eqn:defI}) and the
dilatancy law (Eqn.~\eqref{eqn:nu(i)}), the pressure may be
expressed as a function of the shear rate and the solid fraction :

\begin{eqnarray}
\label{eqn:Plaw}
    P (\nu, \dot\gamma) & = &
    \frac{a^2}{(\nu_{max}-\nu)^{2}} m \dot{\gamma}^{2}.
\end{eqnarray}

\begin{figure}[!htb]
        \begin{center}
            \includegraphics*[width=8cm]{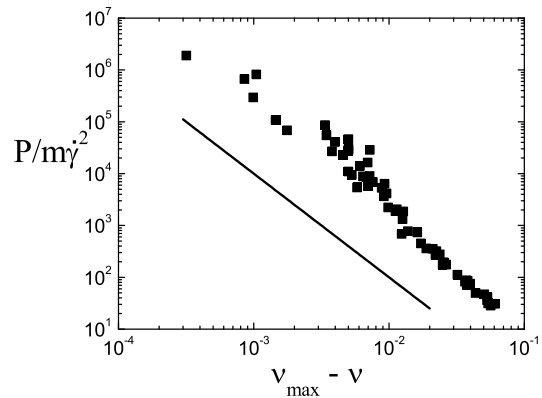}
            \caption{\label{Fig5}
            \textit{$P/\dot{\gamma}^{2}$ as a function of
            ($\nu_{max}-\nu$). The straight line indicates a slope
            of $-2$ ($\mu=0.4$, various $e$).}}
        \end{center}
    \end{figure}

\noindent The divergency with the solid fraction is shown in
Fig.~\ref{Fig5}.

        \subsubsection{Shear stress}
        \label{sec:shearstress}

From the definition of $I$ (Eqn.~\eqref{eqn:defI}) and the
friction law (Eqn.~\eqref{eqn:mu(i)}), the shear stress may be
expressed as a function of the shear rate and the pressure :

\begin{eqnarray} \label{eqn:S(P,dotgamma)}
    S(P,\dot\gamma) & = & \mu^{*}_{min}P + b \sqrt {mP} \dot{\gamma}.
\end{eqnarray}

Using the previous expression of $P$ (Eqn.~\eqref{eqn:Plaw}), it
is also possible to express the shear stress as a function of the
shear rate and the solid fraction :

\begin{eqnarray}
    S(P, \nu, \dot \gamma) & = & \mu^{*}_{min}P +
    \frac{ab}{(\nu_{max}-\nu)} m \dot{\gamma}^{2},
\end{eqnarray}

\noindent or, eliminating $P$ :

\begin{eqnarray} \label{eqn:Slaw}
    S(\nu,\dot \gamma) & =  \frac{ab(\nu^{*} - \nu)}{(\nu_{max}-\nu)^{2}}
    m \dot{\gamma}^{2},
\end{eqnarray}

\noindent with the solid fraction $\nu^{*} = \nu_{max} + a
\mu^{*}_{min}/b$ ($\nu^* \approx 0.86$ for $\mu = 0.4$).

    \subsubsection{Comments}

We first notice that the dilatancy and friction laws, measured in
the whole range of regimes, from the quasi-static to the dynamic,
make the link between the known results in the two extreme regimes
recalled in Sec.~\ref{sec:intro}. In the quasi-static regime,
$\mu^{*}_{min}$ and $\nu_{max}$ may be identified with the
internal friction $\tan \phi$ and the critical solid fraction
$\nu_{c}$ in the critical state (see Sec~\ref{sec:qsreg}).

In the intermediate regime ($I < 0.2$), the expression of the
stress components is analogous to the expression for a homogeneous
system in the collisional regime (Eqs.~\eqref{eqn:kintheo3}). The
dependence on the square of the shear rate, similar to the
original conclusions of Bagnold for concentrated
suspensions~\cite{Bagnold54}, is a consequence of dimensional
analysis~\cite{Lois05}. The dependences on solid fraction are
described by the following dimensionless functions :

\begin{eqnarray}
\left \{ \begin{array}{cccc}
G_P(\nu) =  \frac{a^2}{(\nu_{max}-\nu)^{2}}, \\
\\
G_S(\nu) =  \frac{ab(\nu^{*} - \nu)}{(\nu_{max}-\nu)^{2}}.
\end{array} \right.
\end{eqnarray}

We recall (see Sec.~\ref{sec:collreg}) that the kinetic theory
predicts a divergency in $1/(1-\nu)^2$ from the asymptotic
behavior of the pair correlation function. However, when $\nu \ge
0.67$ (so-called gel transition), then starts a regime of multiple
collisions, strongly correlated, and the divergency rather seems
in $1/(\nu_{max}-\nu)$ ~\cite{Luding01b, Volfson03}. When the
material is sheared, under the effect of cooperative
rearrangements of caged grains, an even stronger divergency of the
viscosity has been conjectured~\cite{Bocquet02b}. It seems that
the precise form of these divergencies are decisive to describe
the shape of the velocity profiles and for the jamming
process~\cite{Bocquet02b, Bocquet02c}. Our quantitative
determination is then a precious information for the modelling of
the dense granular flows.

We also notice that the expression~\eqref{eqn:S(P,dotgamma)} of
$S$ corresponds to a visco-plastic constitutive law, similar to
the ``frictional-collisional" decomposition of the stress tensor,
with a contribution associated to maintained contacts, and a
contribution associated to collisions~\cite{Johnson87a, Mohan97,
Savage98, Ancey01, Volfson03}. In the viscous term, we notice that
the apparent viscosity $b\sqrt{mP}$ is proportional to the square
root of the pressure. The interpretation it that the typical
momentum $m\dot\gamma d$ is exchanged with the inertial time scale
$\sqrt{m/P}$ over a surface of the order $d^2$.

We think that the formulation of the constitutive law through the
dilatancy and friction laws is simpler to use, since it avoids the
treatment of divergency near jamming, which might be a problem in
fluid mechanical numerical simulations.

    \subsection{Influence of the mechanical parameters}
    \label{sec:parametrical}

We shall now describe the influence of the mechanical properties
of the grains ($k_n$, $\mu$ and $e$) on the dilatancy and friction
laws.

\subsubsection{Influence of the elasticity of the grains}

In the limit of rigid grains, that is to say for $\kappa \ge
10^4$, we have not observed any influence of the stiffness
coefficient $k_n$ on the constitutive law. There is however an
influence on the coordination number (see
Sec.~\ref{sec:coordination}). Furthermore, we have shown that
softer grains favor localization of the shear (see
Sec.~\ref{sec:rigid}).

\subsubsection{Influence of the local friction coefficient}

The local friction coefficient $\mu$ has a significant influence
on the dilatancy law. The solid fraction remains a linearly
decreasing function of $I$, but both parameters $\nu_{max}$ and
$a$ depend on $\mu$.

    \begin{figure}[!htb]
        \begin{center}
            \includegraphics*[width=8cm]{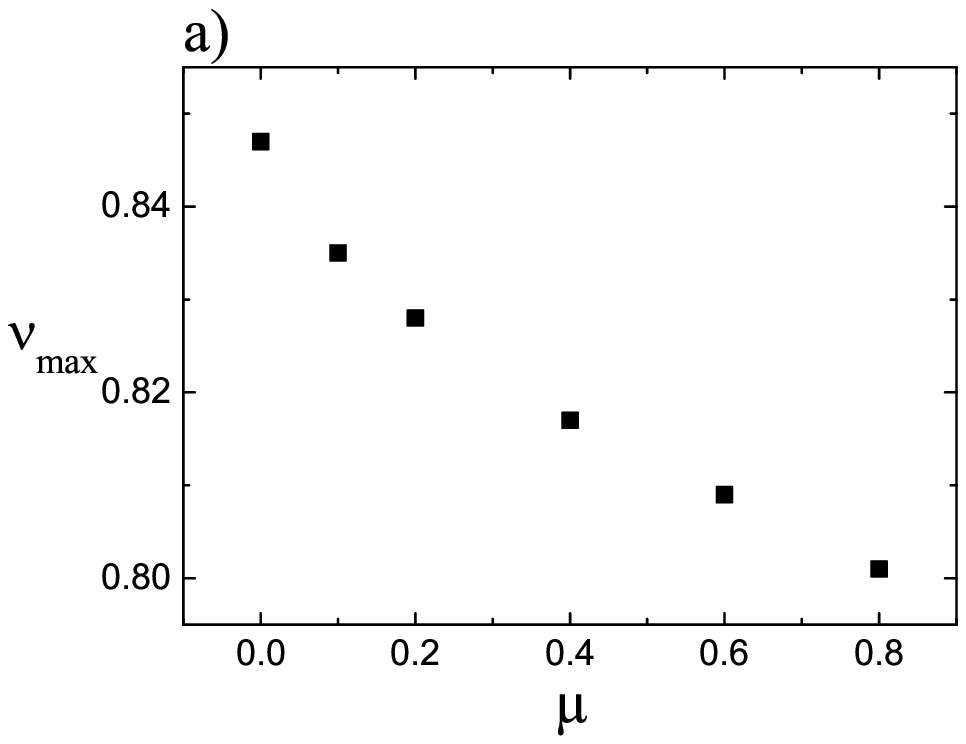}
            \includegraphics*[width=8cm]{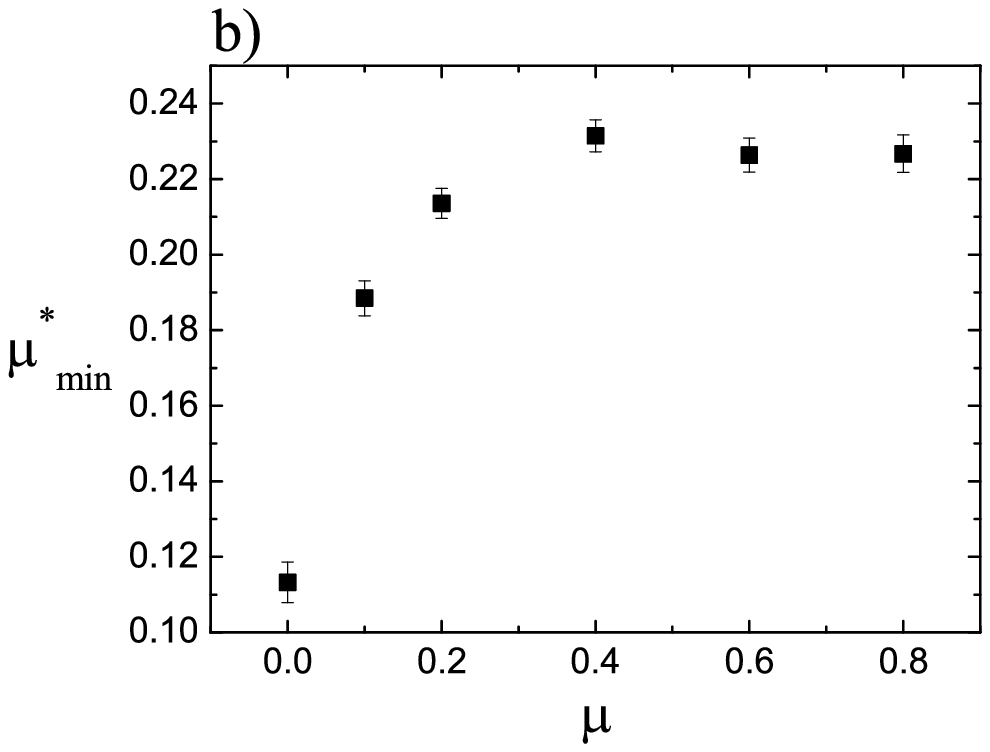}
            \includegraphics*[width=8cm]{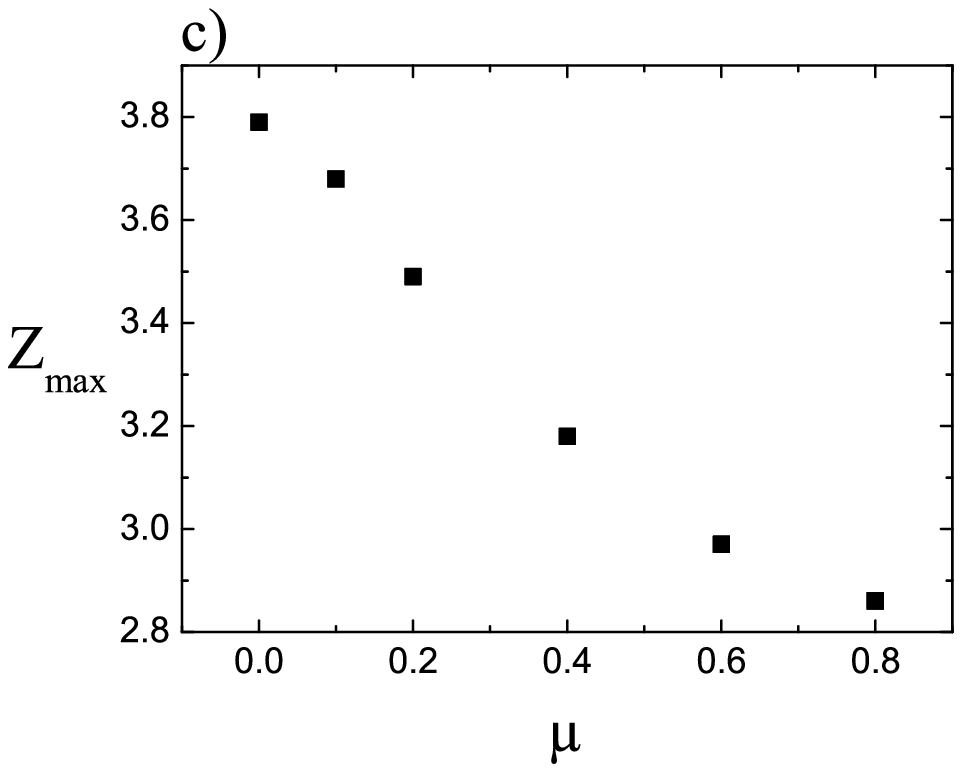}
            \caption{\label{Fig6}
            \textit{Influence of $\mu$ on the critical state ($I=4 \cdot 10^{-3}$ and $e=0.1$):
            (a) Maximum packing fraction $\nu_{max}$,
            (b) Effective friction coefficient $\mu^*_{min}$,
            (c) Coordination number $Z_{max}$.}}
        \end{center}
    \end{figure}

The variation of $a$ is not simple: we measure $a = 0.38$ for $\mu
= 0$, $0.31$ for $\mu = 0.4$ and $0.37$ for $\mu =
0.8$~\cite{Dacruz04a}. Fig.~\ref{Fig6} (a) indicates that
$\nu_{max}$ is a decreasing function of $\mu$ (it is not purely
geometrical in the quasi-static regime). This shows that the solid
fraction, from the critical state to the collisional regime,
depends on the frictional properties of the material. We notice
that our measurements are in agreement with other observations
~\cite{Aharonov99, Campbell02, Roux04b}.

The influence of $\mu$ on the friction law is less significant,
except for frictionless grains ($\mu=0$). The Fig.~\ref{Fig4},
where $\mu$ varies between $0.1$ and $0.8$, shows that $\mu$ has
nearly no influence on $\mu^{*}$ in this range. This variation is
more significant for small $I$. The Fig.~\ref{Fig6} (b) shows more
precisely the variation of the effective friction as a function of
$\mu$ in the quasi-static regime. There is strong variation
between $\mu = 0$ and $\mu = 0.4$, but above $\mu = 0.4$ the
effective friction remains constant.

   \begin{figure}[!htb]
        \begin{center}
            \includegraphics*[width=8cm]{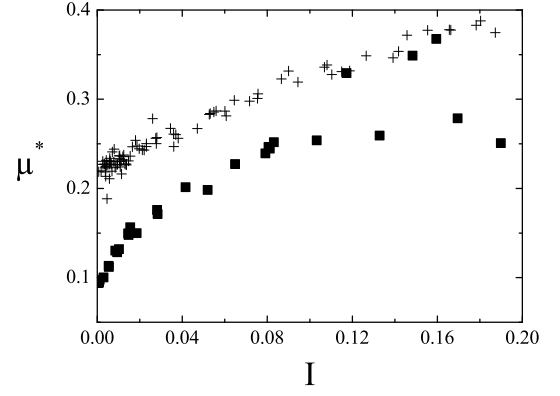}
            \caption{\label{Fig7}
            \textit{Influence of $\mu$ on the friction law ($\mu = 0$ ($\blacksquare$), $\mu \ne 0$ ($+$).}}
        \end{center}
    \end{figure}

As a conclusion, in the case of frictionless grains, the friction
law keeps the same tendency but is shifted toward smaller values
of friction (Fig.~\ref{Fig7}). However, we observe a saturation
for $I \ge 0.1$ if $e = 0.9$, which will be discussed in
Sec.~\ref{sec:collimit}. For $I \le 0.1$, the linear approximation
(Eqn.~\eqref{eqn:mu(i)}) is in trouble : it is rather a sub-linear
dependency, and $\mu^*_{min} \simeq 0.11$. So the case of
frictionless grains is singular and deserves a specific treatment.

    \begin{figure}[!htb]
        \begin{center}
            \includegraphics*[width=8cm]{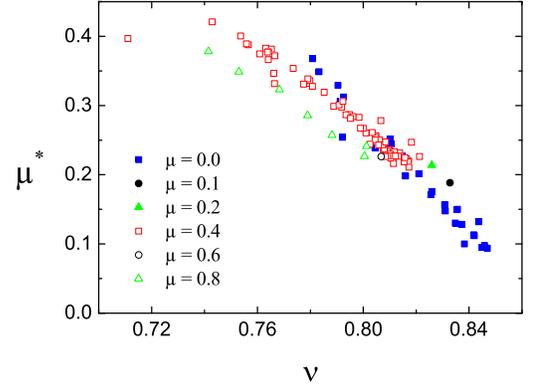}
            \caption{\label{Fig8}
            \textit{(Color online) Variation of the effective friction coefficient as a function of solid fraction.}}
        \end{center}
    \end{figure}

Starting from both variations of solid fraction and effective
friction as a function of the local friction coefficient $\mu$, it
is tempting to draw the variations of the effective friction as a
function of solid fraction instead of inertial number. This is
done on Fig.~\ref{Fig8}. As a matter of fact,
Eqs.~\eqref{eqn:Plaw} and~\eqref{eqn:Slaw} predict :

\begin{equation}
    \mu^{*}(\nu) = \frac{b}{a} (\nu^*-\nu),
\end{equation}

\noindent where $\nu^*$ was previously defined. This new
representation of the results evidences a collapse of the data
(even if $a$, $b$, $\mu^{*}_{min}$ and $\nu_{max}$ vary separately
with $\mu$, $\nu^*= \nu_{max} + a \mu^{*}_{min}/b$ seems
approximately constant). It appears that $\mu^*$ becomes nearly
independent of $\mu$. This master curve is made of complementary
zones of high solid fraction for frictionless grains, and smaller
solid fraction for frictional grains. It is noteworthy that a
small variation of solid fraction (of the order of $10\%$) is
enough to induce a variation of effective friction by a factor $4$
! This decrease of the effective friction when the solid fraction
increases is contrary to the observation in static
compaction~\cite{Horvath96}, but was previously observed under
shear~\cite{Craig86}.

\subsubsection{Influence of the restitution coefficient}

\begin{figure}[!htb]

        \begin{center}
            \includegraphics*[width=8cm]{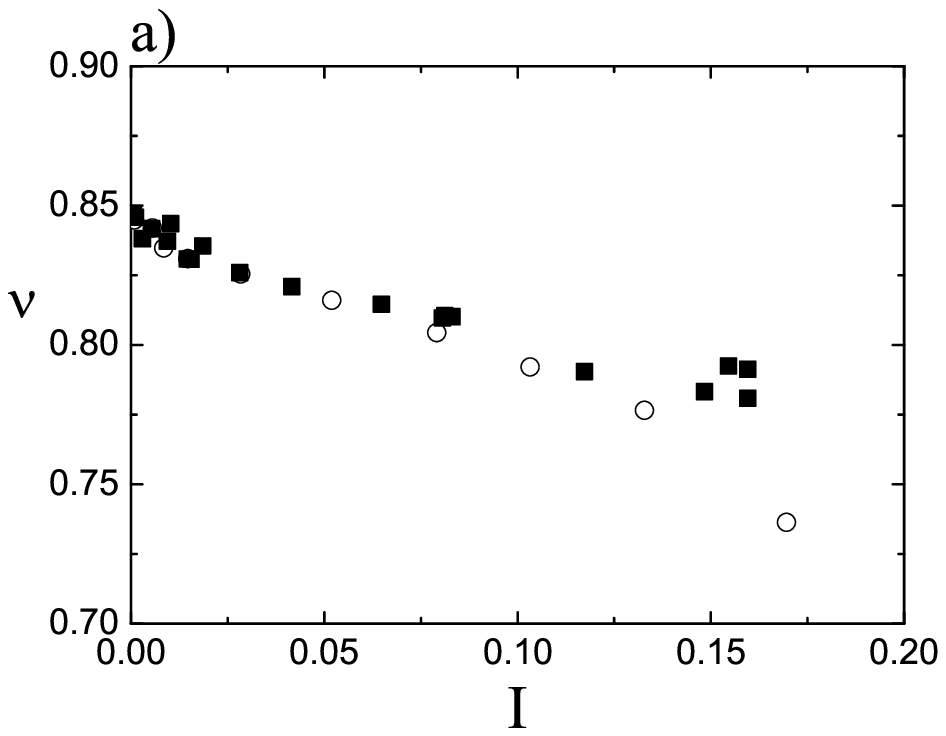}
            \includegraphics*[width=8cm]{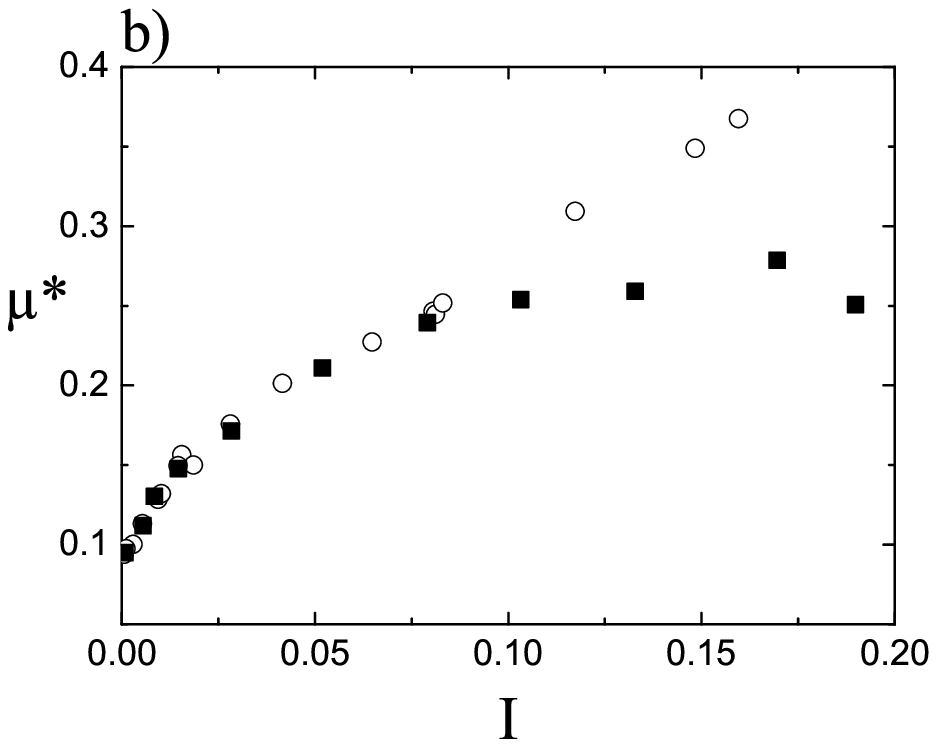}
            \caption{\label{Fig9}
            \textit{Influence of $e$ for $\mu = 0$ ($e = 0.9$ ($\blacksquare$), $e = 0.1$ ($\circ$)) :
(a) Dilatancy law, (b) Friction law.}}
        \end{center}
    \end{figure}

The Fig.~\ref{Fig4} shows that there is no influence of the
restitution coefficient $e$ for frictional grains. For
frictionless grains, the comparison between slightly ($e = 0.1$)
and strongly ($e = 0.1$) dissipative grains reveals
(Fig.~\ref{Fig9}) that there is an influence, limited to the
dynamic regime, for $I \ge 0.1$. Then, as the dissipation
decreases, the dilatancy is less pronounced and the effective
friction saturates (see also~\cite{Campbell02}).

\subsubsection{Collisional limit} \label{sec:collimit}

The situation ($I \ge 0.1 $, $\mu = 0$ and $e = 0.9$) corresponds
to the dense limit of the kinetic theory, with binary
quasi-elastic collisions (then the average contact time tends to
the collision time~\cite{Dacruz04b}). In this dense limit, the
Eqs.~\eqref{eqn:kintheo3} predict a value of the effective
friction independent of $I$ but dependent on $e$, from the values
of the pre-factors $A_i$~\cite{Jenkins85b} :

\begin{equation}
    \mu^{*}(e) = \frac{\sqrt{A_{S} A_{\Gamma}}}{A_{P}} = \frac{1}{2} \left ( \frac{\pi +
    8}{2\pi} \right )^{\frac{1}{2}} \sqrt{1 - e^{2}}.
\end{equation}

For $e = 0.9$, this predicts $\mu^* = 0.29$ which is in fairly
good agreement with the value measured for $I=0.2$ ($0.26$). The
small difference between those two values may be due to the
influence of the walls which induce a sliding velocity and
gradients of the fluctuations~\cite{Hanes88}.

    \section{Fluctuations}
    \label{sec:fluctuation}

Up to now, all the quantities we have studied (solid fraction,
velocities, forces) have been averaged in space and time. In fact,
they are heterogeneous in space and fluctuate in time. The
fluctuations of the forces have already been thoroughly
studied~\cite{Schwarz98, Aharonov99, Howell99b, Radjai01a,
Jalali02}. We have observed~\cite{Dacruz04a} that the fluctuations
of the forces are of the order of a few percents in the
quasi-static regime, increase up to a factor $3$ to $4$ in the
dynamic regime, and that they are larger near the walls than
inside the sheared layer. We shall now discuss in detail the
fluctuations of the motion of the grains.

\subsection{Intermittencies in the quasi-static regime} \label{sec:intermit}

Various studies have shown that granular flows become unstable in
the quasi-static regime. When the velocity is prescribed, one goes
from a continuous flow regime to stick-slip~\cite{Nasuno98,
Lubert01, Ovarlez03}. When the shear stress is prescribed, one
observes an hysteretic and abrupt fluid-solid
transition~\cite{Dacruz02}. We also notice that the observation of
shear localization in discrete simulations of plane shear without
gravity corresponds to the quasi-static regime~\cite{Aharonov02,
Morgan99}. Space-time correlations of the motion of the grains
have been observed~\cite{Roux02, Radjai02, Mueth03, Bonamy02b,
Ferguson03, Chambon03, Pouliquen04}. Several rheological models
have been proposed to describe those observations~\cite{Mills99,
Pouliquen01, Ertas02, Rajchenbach03, Dacruz04a, Volfson03,
Bazant03, Gdr04, Mills05, Lois05}: transmission of forces at the
scale of correlated clusters, two-phase fluid model with order
parameter, activation of rearrangements through the fluctuations
of velocity or forces.

\begin{figure}[!htb]
        \begin{center}
            \includegraphics*[width=8cm]{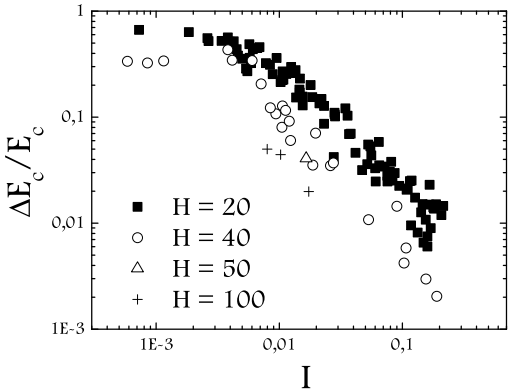}
            \caption{\label{Fig10}
            \textit{Relative fluctuations of the kinetic energy as
            a function of $I$. Influence of $H$.}}
        \end{center}
    \end{figure}

Our study (see~\cite{Dacruz04a} for more details) shows that the
flow is continuous in the intermediate regime, but becomes
intermittent in the quasi-static regime (for $I \le I_{0} \simeq
0.003$). Then the time averaged shear is homogeneous, but the
instantaneous velocity profiles show that the layer oscillates
between two states : a first one where the whole layer is at rest,
caught to the fixed wall, except for a thin layer swept along by
the moving wall, and a second one where the whole layer is swept
along by the moving wall except for a thin layer caught to the
fixed wall. Those two extreme states have a very short duration,
and the system is most of the time in an intermediate state, where
the shear is approximately homogeneous in the whole layer. The
pictures of the fields of velocity and velocity fluctuations show
that the spatial correlations have a vortex like shape with a size
between $3$ and $4$ grain diameter and are of a short duration.
Consequently, the total kinetic energy fluctuates in time, with
sudden peaks associated to the two extremal states, and slow
variations associated to the intermediate situations. The
Fig.~\ref{Fig10} indicate that the relative fluctuations of the
kinetic energy increase when $I$ decreases and saturate in the
quasi-static regime, according to :

\begin{eqnarray}
\label{eqn:Ec}
    \frac{\Delta E_{c}}{<E_{c}>} \simeq \left\{ \begin{array}{cc}
        \frac{c}{I_{0}} & \textrm{for $I < I_{0}$}, \\
        \frac{c}{I} & \textrm{for $I > I_{0}$},
        \end{array} \right.
\end{eqnarray}

\noindent with $I_0 \simeq 0.003$.

\subsection{Velocity fluctuations}

We now discuss the fluctuations of the translation velocity
$\delta v$ and of the rotation velocity $\delta \omega$ of the
grains, which measure the agitation of the granular material. They
are defined by :

\begin{eqnarray}
\left \{ \begin{array}{cccc}
\delta v(y) & = & \sqrt{<\vec v (y)^{2}> - <\vec v (y)>^{2}}, \\
\\
\delta \omega(y) & = &  \sqrt{<\omega (y)^{2}> - <\omega
(y)>^{2}}.
\end{array} \right.\label{eqn:deffluct}
\end{eqnarray}

They are a natural variable in the description of collisional
granular flows (see Sec.~\ref{sec:collreg}). Their influence is
more difficult to analyze in the case of dense flows, where the
motions of the grains are strongly correlated~\cite{Lois05}. The
definition itself of those fluctuations is a problem, since it has
been shown that they depend on the averaging time scale in the
quasi-static regime~\cite{Radjai02}. Our analysis (long time
scale) takes into account both the small fluctuations around the
mean motion (in the ``cage" formed by the nearest
neighbors~\cite{Pouliquen03}), and the large fluctuations
associated to collective motions in the quasi-static regime.

We notice that the quantities defined by Eqs.~\eqref{eqn:deffluct}
are local, \emph{i.e.} function of $y$. Those velocity
fluctuations are uniform at the center of the sheared layer, but
increase near the wall~\cite{Dacruz04a, Dacruz04b}. According to
various models, this agitation of the granular material near the
rough walls is responsible for the increase of the shear rate
(observed in Fig.~\ref{Fig2} (b))~\cite{Denniston99, Bocquet02b,
Prochnow02, Mueth03, Lois05}. In the following, we consider the
average velocity fluctuations in the central part of the sheared
layer, excluding the $5$ first layers near the walls.

    \begin{figure}[!htb]
        \begin{center}
            \includegraphics*[width=8cm]{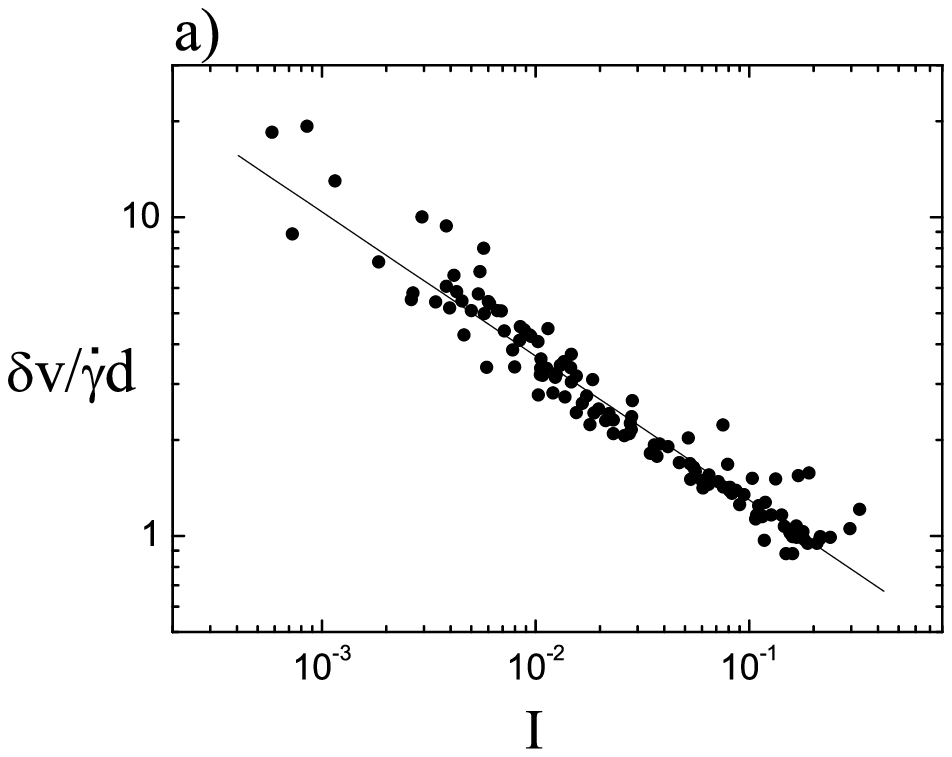}
            \includegraphics*[width=8cm]{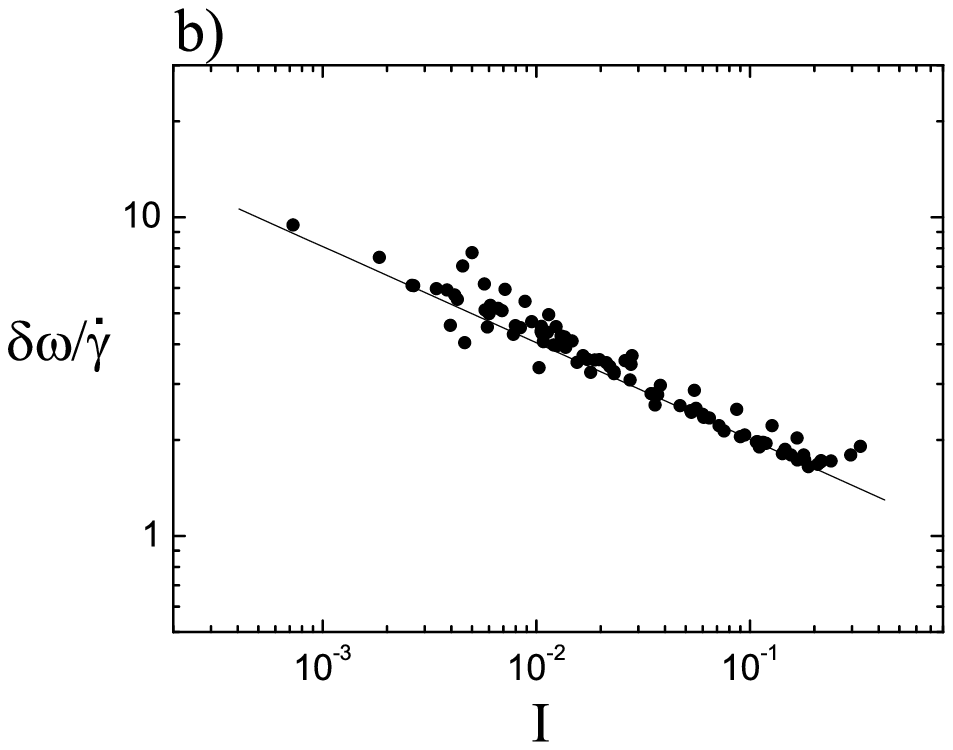}
            \caption{\label{Fig11}
            \textit{Relative velocity fluctuations as
            a function of $I$ :
            (a) Translation velocity, (b) Rotation velocity
            (various parameters).}}
        \end{center}
    \end{figure}

Dimensional analysis suggests to analyze the variations  of the
dimensionless quantities $\delta v/\dot{\gamma}d$ and $\delta
\omega/\dot{\gamma}$ as functions of $I$. We consider
$\dot{\gamma}d$ as the natural scale of translation velocity, and
$\dot{\gamma}$ as the natural scale of rotation velocity (since
$\omega = - \frac{1}{2} \dot{\gamma}$, see
Sec.~\ref{sec:nuvprofiles}). Those variations, drawn on
Fig.~\ref{Fig11}, evidence two scaling laws, independent of the
parameters of the system :

\begin{equation}
    \label{eqn:fluct}
    \left \{ \begin{array}{cccc}
    \frac{\delta v}{\dot{\gamma}d} \simeq \frac{1}{3} I^{-\alpha},\\
    \\
    \frac{\delta \omega}{\dot{\gamma}} \simeq I^{-\beta},
    \end{array} \right.
\end{equation}

\noindent with $\alpha \simeq \frac{1}{2}$ and $\beta \simeq
\frac{1}{3}$, to which we shall refer as translation and rotation
scaling laws.

The velocity fluctuations are significant (larger than $1$) in the
intermediate regime and very significant (larger than $10$) in the
quasi-static regime. In comparison, kinetic theory (see Eqs.
\eqref{eqn:kintheo1} and \eqref{eqn:kintheo2}) predicts that
$\delta v / \dot{\gamma}d = \sqrt{A_s/A_{\Gamma}}$ that is to say
$\sqrt{\frac{\pi+8}{32(1-e^2)}} \simeq 1.3$, for $e=0.9$, which
corresponds to the order of magnitude which is measured in the
dynamic regime.

Large values of $\delta \omega/\dot{\gamma}$ have been observed
experimentally in the quasi-static regime~\cite{Calvetti97} and in
dense flows down inclined planes~\cite{Prochnow02}, and might be
due to the frustration of the rotation~\cite{Kuhn04}.

We give two interpretations of the value of the exponent $\alpha$.
The first explanation~\cite{Gdr04} consists in analyzing the
motion of one grain as a succession of shear phases of duration
$1/\dot \gamma$ with a velocity $\dot \gamma d$ and of sudden
rearrangements with a velocity $d\sqrt{P/m}$ of duration
$\sqrt{m/P}$. This leads to $\frac{\delta v}{\dot{\gamma}d} \simeq
I^{-1/2} \frac{1-I}{1+I}$. The second explanation relies on an
energetic argument. In homogeneous shear, the work of the shear
stress is balanced by the dissipation rate $S \dot{\gamma} =
\Gamma$. If $\Gamma$ describes the dissipation of the fluctuating
kinetic energy $m\delta v^2/2$ during the inertial time
$\sqrt{m/P}$, we obtain : $\delta v/\dot{\gamma}d \simeq
\sqrt{2\mu^*(I)} I^{-1/2}$. For both interpretations, the order of
magnitude of the pre-factor is consistent with the observation.

We now show that the translation scaling law is consistent with
the variations of the relative fluctuations of the kinetic energy
(Eqn.~\eqref{eqn:Ec}). Let us call $\rho=\rho_g \nu$ the average
solid fraction of the granular materials and $\langle E_{c}
\rangle$ the average kinetic energy by unit length, which is
dominated by the translational part : $\langle E_{c} \rangle =
\frac{\rho}{2} \int_{0}^{H} (\dot{\gamma}y)^{2} dy \sim (\rho
V^{2} H)$. In the quasi-static regime, where the system oscillates
between two localized flows, $\Delta E_{c} \sim (\rho V^{2} H)/2$,
so that $\Delta E_{c}/\langle E_{c} \rangle \sim 1$. In the
dynamic regime, $\Delta E_{c} \simeq \frac{\rho}{2} \int_{0}^{H}
\delta v^{2} dy$. Using the translation scaling law for this last
quantity, we get : $\Delta E_{c} \sim \rho V^{2}/(H I)$, so that
$\Delta E_{c}/\langle E_{c} \rangle \sim 1/(H^{2} I)$. This is in
agreement with the dependencies on $I$ and $H$ observed in
Fig.~\ref{Fig10}.

Furthermore, the translation scaling law provides an estimation of
the Reynolds contribution to the stress tensor
(see~Eqn.~\eqref{eqn:sigma}): $\Sigma^{f}/\Sigma \simeq (2\nu/9
\pi) I$. This shows that in the intermediate regime ($I \le 0.1$),
this contribution remains smaller than $1\%$, so that the
contribution of the contact forces $\Sigma^{c}$ remains dominant.

\subsection{Consequences for the constitutive law}

The translation scaling law may also be written :

\begin{equation}
\frac{\delta v}{d} \simeq \frac{1}{3} \dot{\gamma}^{1/2}
(P/m)^{1/4}.
\end{equation}

Consequently, when the pressure is prescribed, the velocity
fluctuations vary like $\dot{\gamma}^{1/2}$, instead of
$\dot{\gamma}$. In the annular shear geometry, the pressure is
constant along the radial direction, and an exponent close to
$1/2$ has been measured experimentally~\cite{Bocquet02b, Mueth03}.

If we introduce the velocity fluctuations in the constitutive law,
like in the collisional regime (see Eqn.~\eqref{eqn:kintheo1}), we
obtain :

\begin{eqnarray}
\left \{ \begin{array}{cccc}
P &\simeq& \frac{9a}{\nu_{max}-\nu} m (\delta v/d)^2,  \\
\\
S &\simeq& \frac{3b\sqrt{a}(\nu^*-\nu)}{(\nu_{max}-\nu)^{3/2}} m
(\delta v/d) \dot{\gamma}.
\end{array} \right.
\end{eqnarray}

Within this formulation, we notice a stronger divergency of the
viscosity near the maximum solid fraction, like in the model
inspired by the glassy dynamics~\cite{Bocquet02c}.

    \section{Contact network}
    \label{sec:network}

In addition to the macroscopic quantities which have been
discussed, the discrete simulations provide information on the
contact network. Its strongly heterogeneous character both in
space and time has already been discussed in detail~\cite{Babic90,
Schwarz98, Aharonov99, Denniston99, Howell99b, Schollmann99,
Latzel00, Radjai01a, Radjai01b, OHern01, Aharonov02}: the contact
time varies from the short collision time in the dynamic regime to
the much longer shear time in the quasi-static
regime~\cite{Zhang00a}, while the distribution of the force
intensity is very wide. We shall not discuss in the following
those distributions of contact time and force intensities. We
shall rather focus on the following three quantities :
coordination number, mobilization of friction and anisotropy of
the contact forces.

        \subsection{Coordination number}
        \label{sec:coordination}

\begin{figure}[!htb]
        \begin{center}
            \includegraphics*[width=8cm]{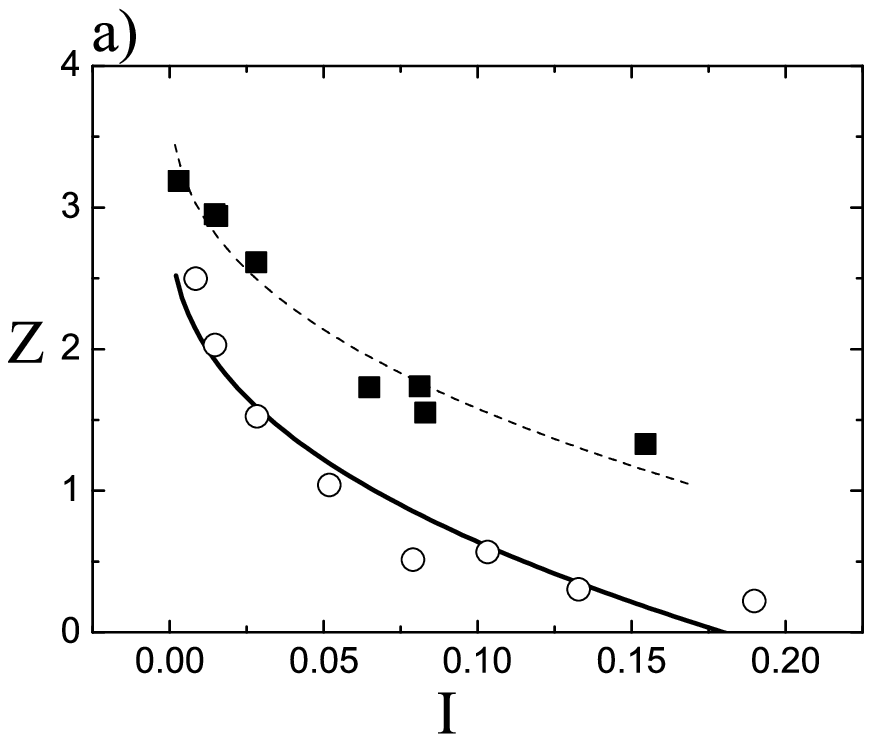}
            \includegraphics*[width=8cm]{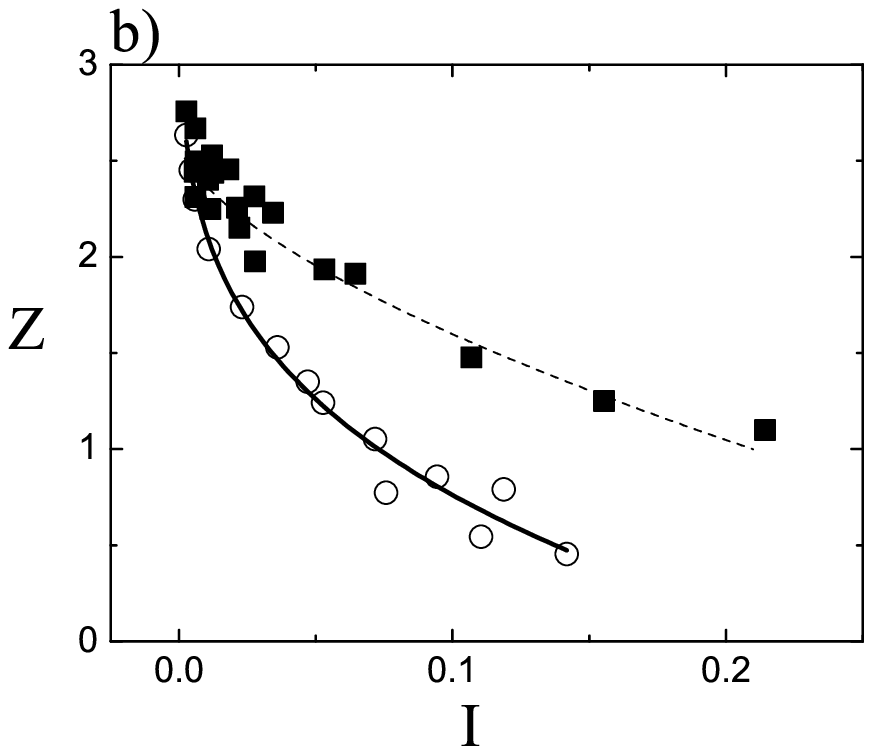}
            \caption{\label{Fig12}
            \textit{Variation of the coordination number $Z$ as a function of $I$
            ($\kappa = 10^{4}$, $e = 0.1$ ($\blacksquare$), $e=0.9$
            ($\circ
            $)) (a) $\mu = 0$ - Fits
            $Z = 3.75 - 7.11 I^{0.51}$ ($- -$),
            $Z = 3.00 - 6.04 I^{0.41}$ ($-$)
            (b) $\mu = 0.4$ - Fits
            $Z = 2.84 - 3.80 I^{0.48}$ ($- -$),
            $Z = 2.60 - 5.48 I^{0.50}$ ($-$).}}
        \end{center}
    \end{figure}

As was shown on Fig.~\ref{Fig1}, the contact network is very
sensitive to the inertial number. A small dilation of the material
(around $10\%$) is enough to observe a transition from a dense
contact network to multiple, or even binary, collisions between
grains. A quantitative indicator is the coordination number $Z$,
that is to say the average number of contacts per grain. The
variations of $Z$ as a function of $I$ are shown in
Fig.~\ref{Fig12}. $Z$ increases as $I$ decreases, and tends toward
a maximum value $Z_{max}$ when $I \to 0$. A possible fit is:

\begin{equation}
    \label{eqn:z(I)}
    Z = Z_{max} - c I^{\gamma},
\end{equation}

\noindent which is drawn on Fig.~\ref{Fig12}, which gathers the
results for a given rigidity number $\kappa$ and various $e$ and
$\mu$.

\begin{figure}[!htb]
        \begin{center}
            \includegraphics*[width=8cm]{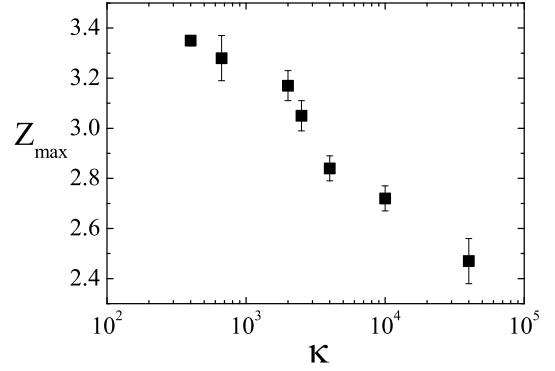}
            \caption{\label{Fig13}
            \textit{Variation of the maximum coordination number as a function of
            $\kappa$ ($e = 0.1$, $\mu = 0.4$).}}
        \end{center}
    \end{figure}

Fig.~\ref{Fig12} shows that the coordination number does not
depend only on the geometry, through the solid fraction, but also
on the mechanical properties of the grains $e$ and $\mu$. The
exponent $\gamma$ is nearly constant ($\gamma \simeq
\frac{1}{2}$), but $Z_{max}$ and $c$ depend on $e$ and $\mu$. When
$\mu$ decreases, $Z$ increases and $Z_{max}$ tends to $4$ for
frictionless grains~\cite{Roux00} (see Fig.~\ref{Fig6} (c)). We
also notice that $Z$ increases when $e$ decreases, due to the
increasing collision time. The Fig.~\ref{Fig13} indicates that
$Z_{max}$ decreases significantly with $\kappa$, as expected.
Among the various quantities which we have studied, the
coordination number is the only one which varies significantly
with the rigidity of the grains.

        \subsection{Mobilization of friction}

\begin{figure}[!htb]
        \begin{center}
            \includegraphics*[width=8cm]{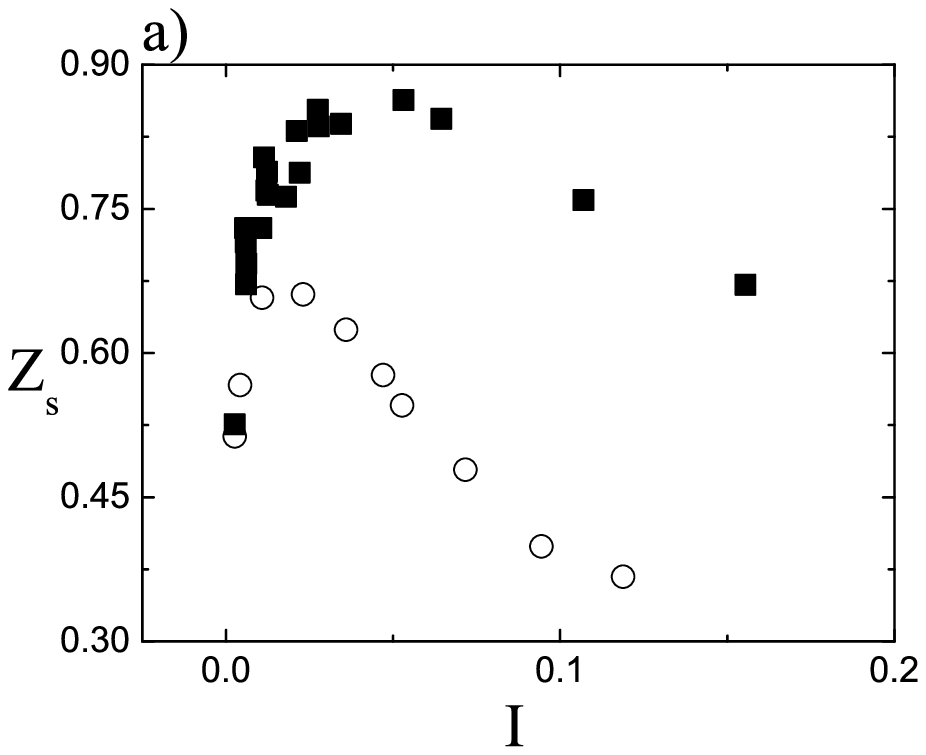}
            \includegraphics*[width=8cm]{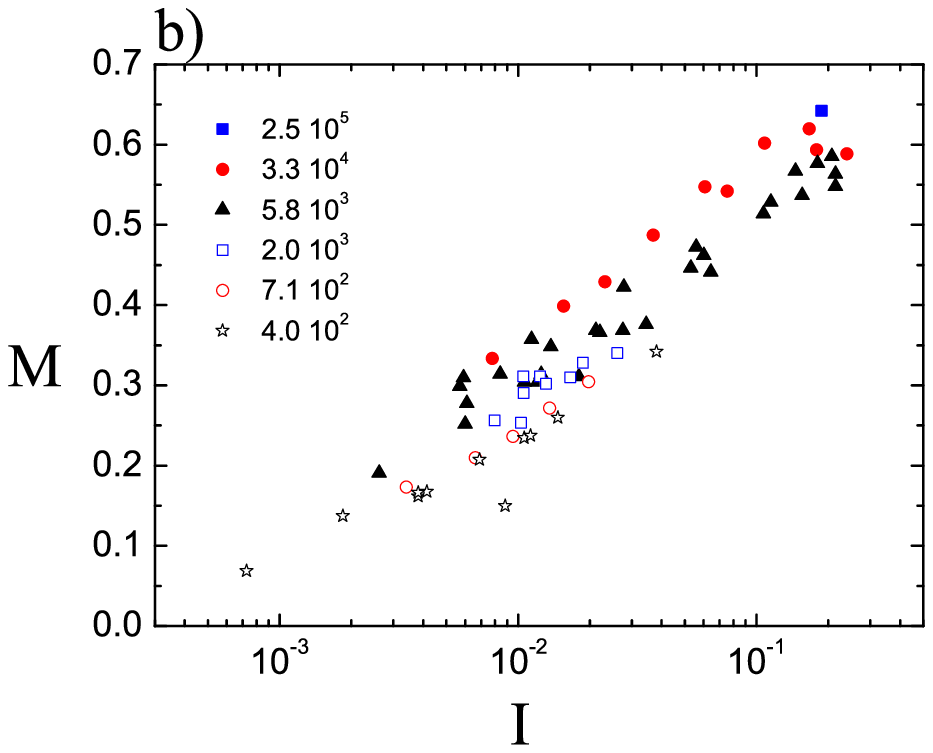}
            \caption{\label{Fig14}
            \textit{(Color online) (a) Variation of $Z_s$ as a function of
            $I$ ($\mu = 0.4$ - $\kappa = 4000$ - $e=0.1$ ($\blacksquare$), $e=0.9$ ($\circ$))
            (b) Variation of $M$ as a function of $I$,
            for various $\kappa$.}}
        \end{center}
    \end{figure}

Inside the population of contacts, and for frictional grains, we
now introduce a distinction between the ``sliding" contacts where
the local friction is completely mobilized ($\vert T \vert = \mu
N$) and the other ``rolling" contacts ($\vert T \vert \le \mu N$).
This distinction is slightly different from the one proposed
in~\cite{Volfson03}, which distinguishes ``fluid" contacts
(collisions and sliding enduring contacts ) and ``solid" contacts
(rolling enduring contacts). We define $Z_{s}$ as the average
number of sliding contacts per grain~\cite{Staron02b}. The
Fig.~\ref{Fig14} (a) shows the variations of $Z_s$ as a function
of the inertial number $I$. We observe that $Z_s$ increases with
$I$ in the quasi-static regime, up to a maximum in the
intermediate regime. Moreover, the Fig.~\ref{Fig14} (a) indicates
that the $Z_{s}(I)$ curve depends on the restitution coefficient
$e$.

We have shown on Fig.~\ref{Fig14} (b) the variations with $I$ of
the ratio $M = Z_s/Z$, which, as the proportion of sliding
contacts, is an indicator of the mobilization of friction. We
observe that, contrarily to $Z_s$, $M$ increases, approximately
logarithmically, as a function of $I$. We also notice a slight
increase of $M$ when $\kappa$ increases.

        \subsection{Anisotropy of the contact network}

   \begin{figure}[!htb]
        \begin{center}
            \includegraphics*[width=6cm]{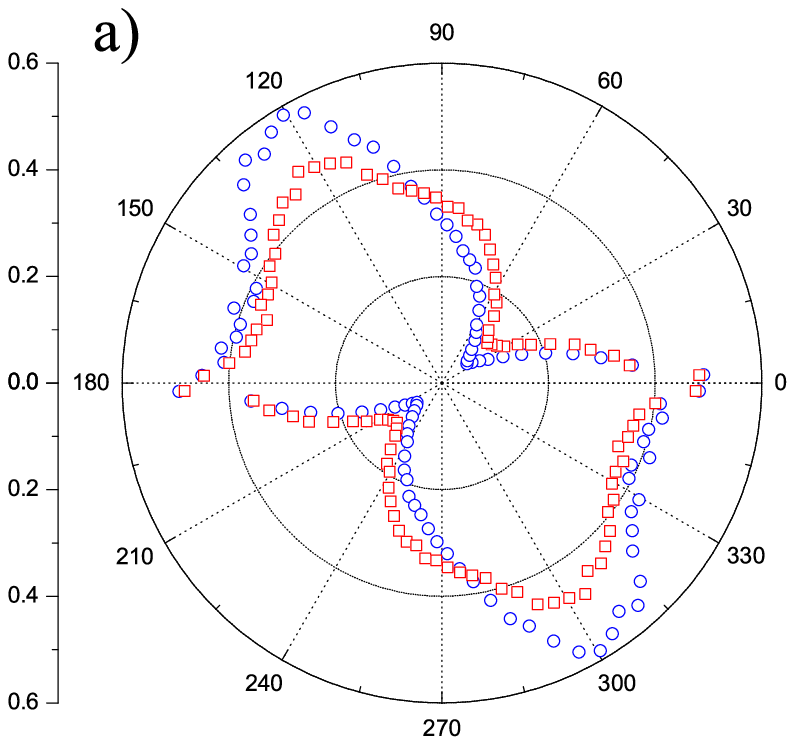}
            \includegraphics*[width=6cm]{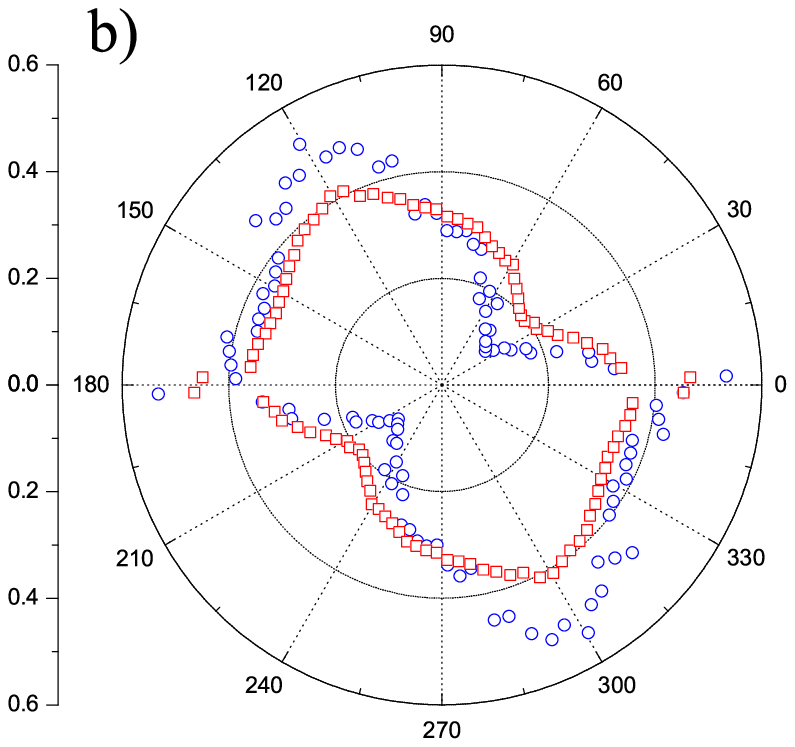}
            \includegraphics*[width=6cm]{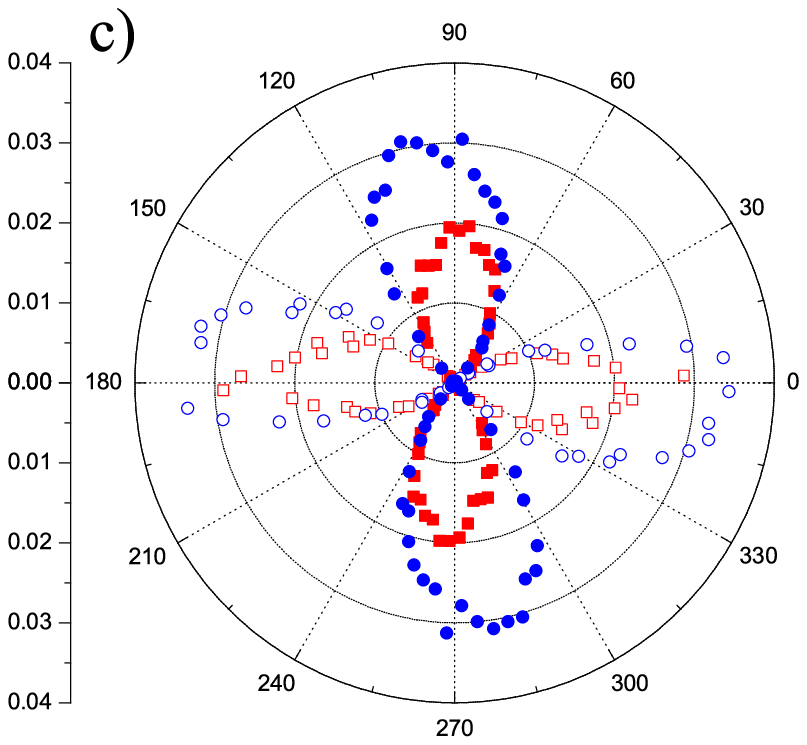}
            \caption{\label{Fig15}
            \textit{(Color online) Angular distribution of the contact forces
            (quasi-static regime ($I = 0.005$ - $\square$), dynamic regime ($I = 0.13$ -
            $\circ$):
            (a) $\zeta_N(\phi)$ ($\mu = 0.4$),
            (b) $\zeta_N(\phi)$ ($\mu = 0$),
            (c) $\zeta_T(\phi)$ ($\mu = 0.4$)
            - $\ge 0$ white symbols, $\le 0$ black symbols.
            }}
        \end{center}
    \end{figure}

We now discuss the anisotropies of the contact network. We call
$\phi$ the direction of a contact counted counterclockwise from
the $x$ direction, between $0$ and $\pi$. ($\vec n_{\phi}, \vec
t_{\phi}$) is the local frame in the direction $\phi$.  Let us
call $\langle N \rangle$ the average normal force in the
homogeneous layer, then $\langle N(\phi) \rangle$ and $\langle
T(\phi) \rangle$ the average normal and tangential forces in the
homogeneous layer in the direction $\phi$. The anisotropies are
described by the three angular distributions of contact
orientations $\rho(\phi)$, of intensities of normal forces
$\xi_N(\phi)= \langle N(\phi) \rangle / \langle N \rangle$ and of
intensities of tangential forces $\xi_T(\phi) = \langle T(\phi)
\rangle / \langle N \rangle$. Then we define $\zeta_N(\phi) =
\rho(\phi) \xi_N(\phi)$ and $\zeta_T(\phi) = \rho(\phi)
\xi_T(\phi)$. Those angular distributions satisfy the
normalization relations :

\begin{eqnarray}
\left \{ \begin{array}{cccc}
\int_0^{\pi} \rho(\phi) d\phi &=& 1,\\
\\
\int_0^{\pi} \zeta_N(\phi) d\phi &=& 1,\\
\\
\int_0^{\pi} \zeta_T(\phi) d\phi &=& 0.
\end{array} \right.
\end{eqnarray}

We show in Fig.~\ref{Fig15} the two quantities $\zeta_N(\phi)$ and
$\zeta_T(\phi)$ which will be useful in the discussion of the
friction law. We distinguish first quasi-static and dynamic
regimes, second frictional and frictionless grains. A positive
value of $\zeta_T(\phi)$ indicates that the tangential forces
induce on average a counterclockwise rotation of the grains, and
is represented with white symbols in Fig.~\ref{Fig15} (c). A
negative value of $\zeta_T(\phi)$ indicates that the tangential
forces induce on average a clockwise rotation of the grains, and
is represented with black symbols in Fig.~\ref{Fig15} (c).

We notice a strong anisotropy of the contact network with
privileged orientations for $\zeta_N(\phi)$ along the directions
of shear ($\phi \simeq 0$ and $\pi$) and of maximum compression
($\phi \simeq 2\pi/3$), and for $\zeta_T(\phi)$ along the
directions of shear ($\phi \simeq 0$ and $\pi$) and of the shear
gradient ($\phi \simeq \pi/2$). Those anisotropies slightly change
between the quasi-static and dynamic regimes and between
frictional and frictionless grains cases.

Those anisotropies may be explained within a very simplified
picture of a granular material organized in layers along the shear
direction. Then, there are two kinds of contacts between grains,
inside a layer ($\phi \simeq 0$ and $\pi$) and between layers
($\pi/3 \le \phi \le 2\pi/3$). The contacts between layers are
created along the direction of maximum compression ($\phi \simeq
2\pi/3$). In the quasi-static regime, those contacts are
maintained up to the point where the grains separate ($\phi \simeq
\pi/3$). This is in contrast with the dynamic regime where the
grains bounce, so that $\zeta_N$ is stronger around $\pi/3$ and
smaller between $\pi/3$ and $2\pi/3$.

\begin{figure}[!htb]
        \begin{center}
            \includegraphics*[width=8cm]{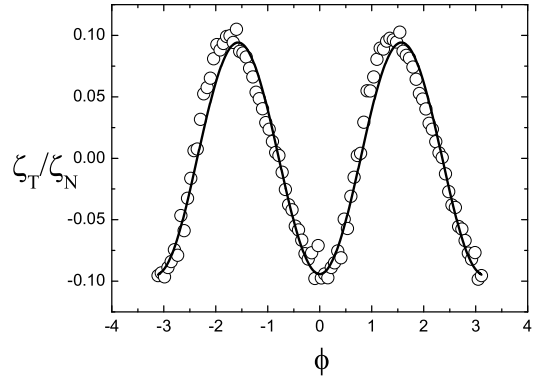}
            \includegraphics*[width=8cm]{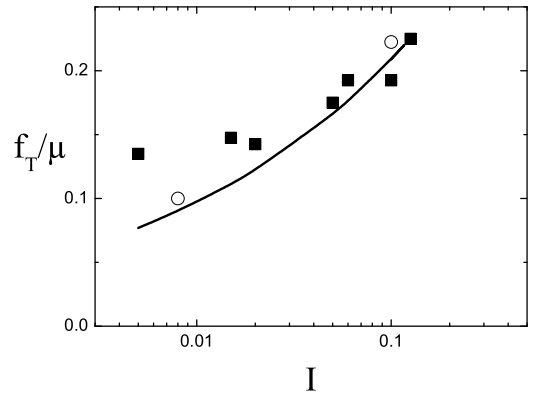}
            \caption{\label{Fig16}
            \textit{Angular distribution of tangential contact
            forces:
            (a) $\zeta_T(\phi)/\zeta_N(\phi)$ ($\circ$) fitted by
            $f_T \cos(2\phi)$ ($-$) for $I = 0.13$ and $\mu = 0.4$,
            (b) $f_T(I)/\mu$ for $\mu = 0.4$ ($\blacksquare$) and $\mu = 0.8$ ($\circ$)
            compared with $\frac{\sqrt{2}}{\pi}I^{1/3}$.
            }}
        \end{center}
    \end{figure}

We observe (see Fig.~\ref{Fig16} (a)) that the tangential
anisotropy is well described by the following expression :

\begin{equation}\label{eqn:zetaT(I)}
\zeta_T(\phi) = f_T(I) \zeta_N(\phi) \cos(2\phi),
\end{equation}

\noindent with an increasing positive function $f_T(I)$. This
means that the contacts between layers favor a clockwise rotation
of the grains, while the contacts inside layers favor a
counterclockwise rotation of the grains. Furthermore, this shows
that the average tangential force is smaller in the quasi-static
regime than in the dynamic regime.

\section{Microscopic origin of the friction law} \label{sec:frictionlaw}

We now try to understand quantitatively the macroscopic friction
law on the basis of the previous microscopic information
(fluctuations and anisotropy of the contact network).

\subsection{Friction and rotation}

We first discuss the tangential anisotropy and show how it is
related to the rotation of the grains~\cite{Kuhn04}.

\subsubsection{Average rotation}

In a first step, we take into account the average rotation
velocity $\omega$. We consider the homogeneous shear of an
assembly of grains of diameter $d$ with an average shear rate
$\dot \gamma$. Then, the tangential relative velocity between two
grains (see Sec.~\ref{sec:contactlaw}) is given by :

\begin{equation}
V^T(\phi)= d (\omega + \dot \gamma \sin^2(\phi)).
\end{equation}

\noindent The average tangential force exerted on a grain should
be equal to zero in steady state :

\begin{equation}
\int_0^{\pi} \rho(\phi) T(\phi) d\phi = 0.
\end{equation}

\noindent We start with two very crude assumptions : all the
contacts are sliding ($T = - \mu N sign(V^T)$) and the normal
force distribution is isotropic ($\zeta_N(\phi) = 1/\pi$).
Consequently :

\begin{equation}
\int_0^{\pi} sign(\omega + \dot \gamma \sin^2(\phi)) d\phi = 0.
\end{equation}

\noindent This is an explanation for the relation $\omega = -
\frac{1}{2} \dot{\gamma}$, which was described in
Sec.~\ref{sec:nuvprofiles} (see Fig.~\ref{Fig2} (b)). Consequently
$V^T(\phi)= -d \dot \gamma \cos(2\phi)/2$, $\zeta_T = \mu/\pi$
when $0 \le \phi \le \pi/4$ and $3\pi/4 \le \phi \le \pi$ and
$\zeta_T = -\mu/\pi$ when $\pi/4 \le \phi \le 3\pi/4$. In
comparison with Fig.~\ref{Fig15} (c), we notice that the sign of
$\zeta_T(\phi)$ is correct, but that the order of magnitude is too
large, by a factor of around $10$.

\subsubsection{Fluctuations of the rotation}

As a way to understand the order of magnitude of $\zeta_T(\phi)$,
we now take into account the fluctuations of the rotation
velocity, which have been evidenced in Sec.~\ref{sec:fluctuation}.
Denoting as $\delta \omega_{i,j}$ the fluctuations of rotation of
two grains $i$ and $j$ in contact, their relative tangential
velocity at the contact point becomes  :

\begin{equation}
V^T_{ij}(\phi)= -d \dot \gamma \cos(2\phi)/2 + d/2 (\delta
\omega_i +\delta \omega_j).
\end{equation}

\noindent Keeping the assumption of sliding contacts, we predict :

\begin{equation}
\zeta_T(\phi) = - \mu \zeta_N(\phi) \langle sign(V^T) \rangle,
\end{equation}

\noindent where $\langle sign(V^T) \rangle$ is the statistical
average over the fluctuating rotations of the two grains. Discrete
numerical simulations have shown that the distribution of the
rotation velocity is approximately lorentzian~\cite{Prochnow02}.
We make the assumption that the fluctuations of rotation of two
grains in contact are not correlated. Then the random variable
$(\delta \omega_i +\delta \omega_j)/2$ follows a lorentzian
distribution, with a zero mean value and a variance $\delta
\omega/\sqrt{2}$. Then we obtain :

\begin{equation}
\langle sign(V^T) \rangle = - \frac{2}{\pi} \arctan \left(
\frac{\dot\gamma}{\sqrt{2}\delta \omega} \cos(2\phi)\right).
\end{equation}

\noindent Using the rotation scaling law~\eqref{eqn:fluct} :

\begin{equation}
\langle sign(V^T) \rangle = - \frac{2}{\pi} \arctan \left(
\frac{I^{1/3}}{\sqrt{2}}\cos(2\phi)\right),
\end{equation}

\noindent and for small $I$ ($\le 0.2$) :

\begin{equation}
\langle sign(V^T) \rangle \simeq - \frac{\sqrt{2}}{\pi} I^{1/3}
\cos(2\phi),
\end{equation}

\noindent so that at the end :

\begin{equation}\label{eqn:zetaT(I)2}
\zeta_T(\phi) \simeq \frac{\sqrt{2}}{\pi} \mu I^{1/3}
\zeta_N(\phi) \cos(2\phi).
\end{equation}

This expression reproduces the observed angular dependence in
$\zeta_N(\phi) \cos(2 \phi)$, shown in Fig.~\ref{Fig16} (a). The
prediction for the dependence of the amplitude on $I$ and $\mu$ is
in agreement with the observed pre-factor $f_T(I)$ defined in
Eqn.~\eqref{eqn:zetaT(I)}, as shown in Fig.~\ref{Fig16} (b). The
order of magnitude is now consistent with the observations.

As a conclusion, the fluctuations of the rotation velocity are a
possible quantitative explanation of the angular distribution of
tangential forces $\zeta_T(\phi)$. When $I$ decreases, the
relative fluctuations of rotation increase, so that the average
relative tangential velocity of two grains in contact tends to
zero, which kills the frictional effect. This model is very crude.
In the dynamic collisional regime, we have seen that most of the
contacts are sliding (see Fig.~\ref{Fig14}) and the assumption of
uncorrelated fluctuations may seem reasonable. On the contrary, in
the quasi-static regime, we have observed that most of the
contacts are rolling. Furthermore, our simulations reveal
correlations of the rotations of grains in contact: it seems that
the flowing granular material is organized in clusters of grains
rotating in the same way~\cite{Kuhn04}. Such correlations of the
grain motion deserves further study~\cite{Pouliquen04, Mills05}.

\subsection{Friction and anisotropy}

We now discuss the friction law $\mu^*(I)$ on the basis of the
information on the contact network. We would like to understand
the increase of $\mu^*$ with $I$, and its dependence with the
microscopic friction $\mu$.

A first possible interpretation lies in the increase of the
mobilization of friction $M(I)$ (Fig.~\ref{Fig14} (b)): most of
the contacts are rolling in the quasi-static regime, while most of
them are inelastic sliding collisions in the dynamic regime.
Consequently, the energy dissipation, and hence the effective
friction, should be stronger in the dynamic regime. However, since
the effective friction coefficient of an assembly of frictionless
grains is not equal to zero, this interpretation is certainly not
sufficient. We are now going to show the crucial role of the
anisotropies of the contact network.

We consider the homogeneous shear of an assembly of grains of
diameter $d$ with average solid fraction $\nu$, coordination
number $Z$, and normal force $\langle N \rangle$. The stress
tensor is dominated by the contribution of the contacts
(see~Eqn.~\ref{eqn:sigma}). It is possible to express it as a
function of the angular distributions of contact forces, which we
have previously defined ~\cite{Kanatani81, Cambou95, Calvetti97,
Prochnow02}:

\begin{equation}
  \underline{\underline{\Sigma}} = \frac{2\nu Z <N>}{\pi d} \int_0^{\pi} [\zeta_N(\phi) \vec
   n_{\phi} + \zeta_T(\phi) \vec t_{\phi}] \otimes \vec
    n_{\phi} d\phi.
\end{equation}

\noindent Using the properties of the stress tensor ($\Sigma_{xy}
= \Sigma_{yx}$, and $\Sigma_{xx} = \Sigma_{yy}$) and the
normalization of angular distributions, the effective friction
coefficient takes the simple expression :

\begin{equation}\label{eqn:prediction-mu*}
\mu^* = - \int_0^{\pi} [\zeta_N(\phi) \sin(2\phi) + \zeta_T(\phi)
\cos(2\phi)] d\phi.
\end{equation}

In the first term associated to the normal forces ($\mu^*_N$), the
factor $\sin(2\phi)$ is positive for $\phi$ between $0$ and
$\pi/2$, so that it decreases the effective friction, and negative
for $\phi$ between $\pi/2$ and $\pi$, so that it increases the
effective friction. Consequently, the evolution of the angular
distribution between the quasi-static regime and the dynamic
regime (see Fig.~\ref{Fig15} (a) and (b)) might explain part of
the increase of the effective friction.

\begin{figure}[!htb]
\begin{center}
            \includegraphics*[width=8cm]{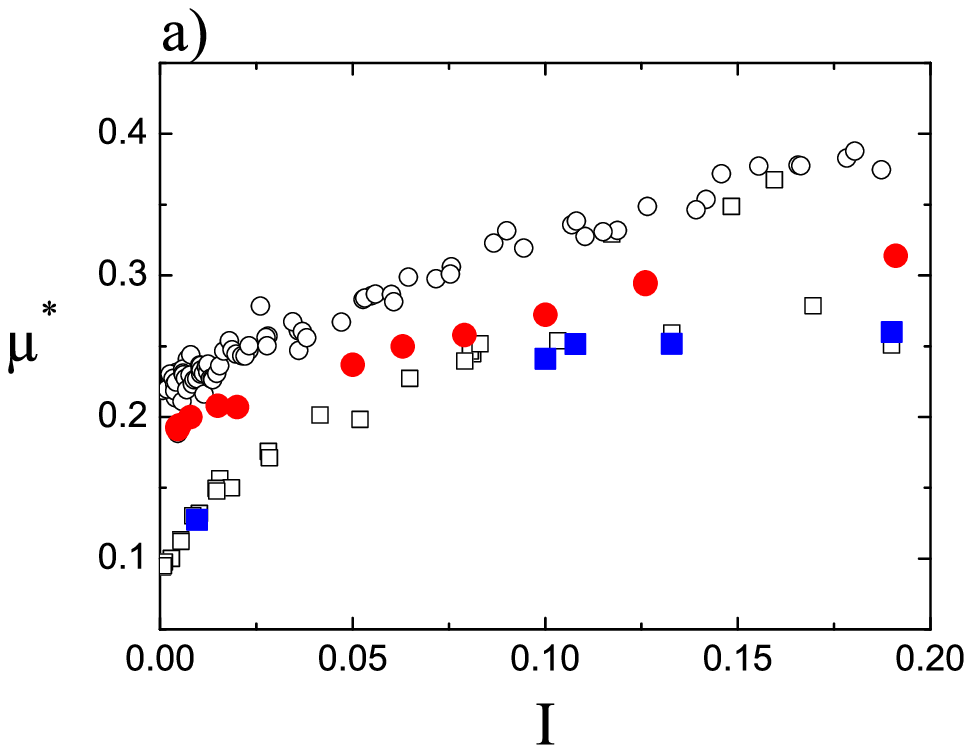}
            \includegraphics*[width=8cm]{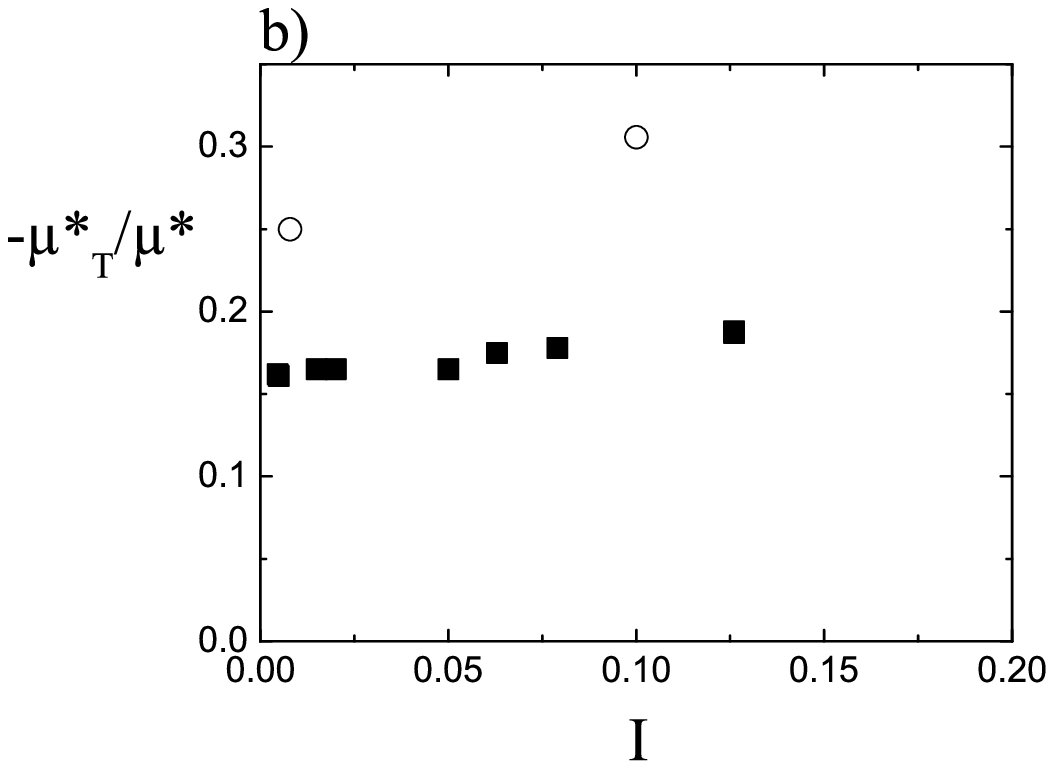}
            \caption{\label{Fig17}
            \textit{(Color online) Friction and anisotropy :
            (a) Comparison of Eqn.~\ref{eqn:prediction-mu*} (black symbols)
             with direct measurement of Fig.~\ref{Fig7} (open symbols)
             ($\mu = 0$ ($\square$), $\mu \ne 0$ ($\circ$)),
             (b) Tangential contribution for $\mu=0.4$ ($\blacksquare$) and
             $\mu=0.8$ ($\circ$).}}
\end{center}
\end{figure}

It is possible to give an estimation of the second term associated
to the tangential forces ($\mu^*_T$), using the approximation
\eqref{eqn:zetaT(I)2} for $\zeta_T(\phi)$, with $\zeta_N(\phi) =
1/\pi$. We obtain $\mu^*_T \simeq -(\mu/\sqrt{2}\pi) I^{1/3}$.
This contribution is negative and small. The complete calculation
(Fig.~\ref{Fig17} (b)) confirms that the contribution of
tangential forces to the effective friction ($\vert \mu^*_T
\vert/\mu*$) remains of the order of $15\%$ for $\mu = 0.4$ and
$30\%$ for $\mu = 0.8$.

Fig.~\ref{Fig17} (a) compares the prediction based on the angular
distribution~\eqref{eqn:prediction-mu*} with the complete
calculation based on~\eqref{eqn:sigma}, both for frictional and
frictionless grains (the results of Fig.~\ref{Fig7}). The
agreement is very good for frictionless grains, but we observe a
deviation for frictional grains, of the order of $15\%$, which we
do not explain at this stage. A possible explanation is the small
polydispersity since the monodispersity was the only simplifying
assumptions leading to~\eqref{eqn:prediction-mu*}.

This result is very paradoxical. Contrarily to our first tentative
explanation of the friction law, rather than explaining the
macroscopic friction, the microscopic friction (tangential forces)
has a small and even negative contribution to the macroscopic
friction. The friction law depends mostly on the angular
distribution of normal forces. A small variation of the
distribution is enough to increase the effective friction by a
factor of two between the quasi-static and the dynamic regime. In
the same way, this distribution is more isotropic in the case of
frictionless grains, and this explains the shift of the friction
law. The microscopic friction has an indirect effect on the
friction law, through the modification of the angular distribution
of normal forces. This would mean that the very origin of the
visco-plastic constitutive law relies in the anisotropy of the
contact network (fabric) in response to the shear. This point
which has already been studied in the quasi-static
regime~\cite{Radjai04} would deserve further study.

\section{Influence of gravity}
\label{sec:gravity}

Up to now, we have studied the simplest flow geometry, plane shear
between two rough walls, without gravity. Then the stress
distribution is homogeneous. In the limit of rigid grains and in
the intermediate regime, we have observed steady homogeneous
flows, in which we have measured the dilatancy and friction laws.
Other simulations of homogeneous shear states without walls have
confirmed those observations~\cite{Rognon05a}.

We now wonder if those laws are valid in other flow geometries
(annular shear, vertical chute, inclined plane...) where, as a
consequence of the heterogeneous stress distribution, the flow is
no more homogeneous, and often localized near the walls or the
free surface~\cite{Gdr04, Dacruz04a}.

As a simple example of an heterogeneous stress distribution, we
still consider the plane shear flow geometry, but we study the
influence of gravity $g$ along $y$ (oriented toward the fixed
wall). Then, the shear stress $S$ remains constant in the sheared
layer, but the pressure varies along $y$ according to :

\begin{eqnarray}
  P(y) &=& P_w + \rho_g g \int_y^H \nu(t) dt.
\end{eqnarray}

\noindent where $P_w$ is the pressure prescribed on the moving
wall. We notice that this kind of flow has been previously studied
experimentally~\cite{Savage84, Hanes85b, Craig86, Nasuno98,
Tsai04}, numerically~\cite{Thompson91, Zhang92, Jalali02,
Volfson03} and theoretically~\cite{Johnson87a}.

\subsection{Dimensional analysis}

We define dimensionless numbers from the various time scales of
the system, and notice that in an heterogeneous system, those
quantities depend on $y$. In addition to the shear time $1/\dot
\gamma(y)$, the inertial time $\sqrt{m/P(y)}$, and the collision
time $\tau_c$, the presence of gravity introduces another gravity
timescale $\sqrt{d/g}$. Consequently, in addition to the inertial
$I(y)$ and rigidity $\kappa(y)$ numbers, we must consider a third
dimensionless number, measuring the intensity of gravity. We call
$G(y)$ this gravity number, defined as the ratio of inertial to
gravity times :

\begin{equation}
G(y) = \sqrt{\frac{gm}{P(y)d}}.
\end{equation}

In the following, we have not studied the influence of the
rigidity of the grains : $\kappa(y)$ varies between $10^4$ near
the moving wall and $10^2$ near the fixed wall, where we are
beyond the rigid grain limit. The simulated systems ($H/d \simeq
40$) are described by the two dimensionless numbers $I_g=V/H
\sqrt{m/P_w}$ and $G=\sqrt{\frac{gm}{P_wd}}$. $I_g$ varies between
$0.0025$ and $0.05$, whereas $G$ varies between $0$ and $3.2$.

\subsection{Localization}

Let us first assume that the dilatancy and friction law identified
in homogeneous shear flows are still valid in presence of gravity.
Then we predict that, like $S/P(y)$, $I(y)$ should decrease from
$y=H$ to $y=0$. Consequently the shear rate should decrease and
the solid fraction increase. For a large enough $G$, there is a
point $(H-\lambda)$ where $S/P(H-\lambda)=\mu^*_{min}$. Below this
point, the granular material should be at rest: $\dot \gamma = 0$
and $\nu=\nu_{max}$. This would mean that the shear is localized
in a layer of width $\lambda$.

 \begin{figure}[!htb]
        \begin{center}
            \includegraphics*[width=7cm]{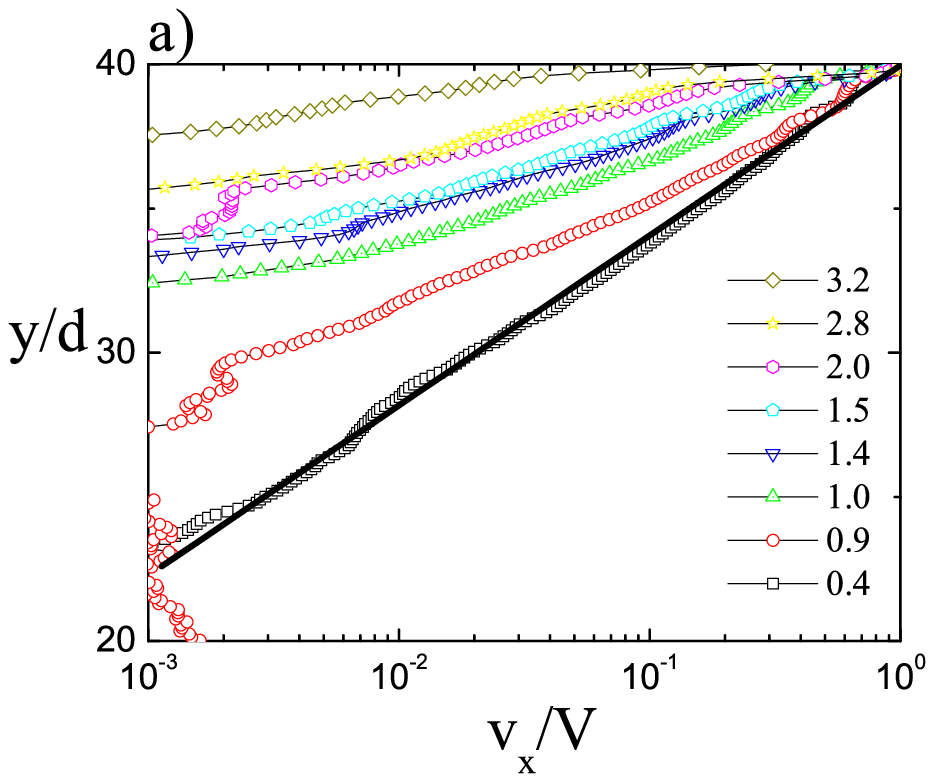}
            \includegraphics*[width=7cm]{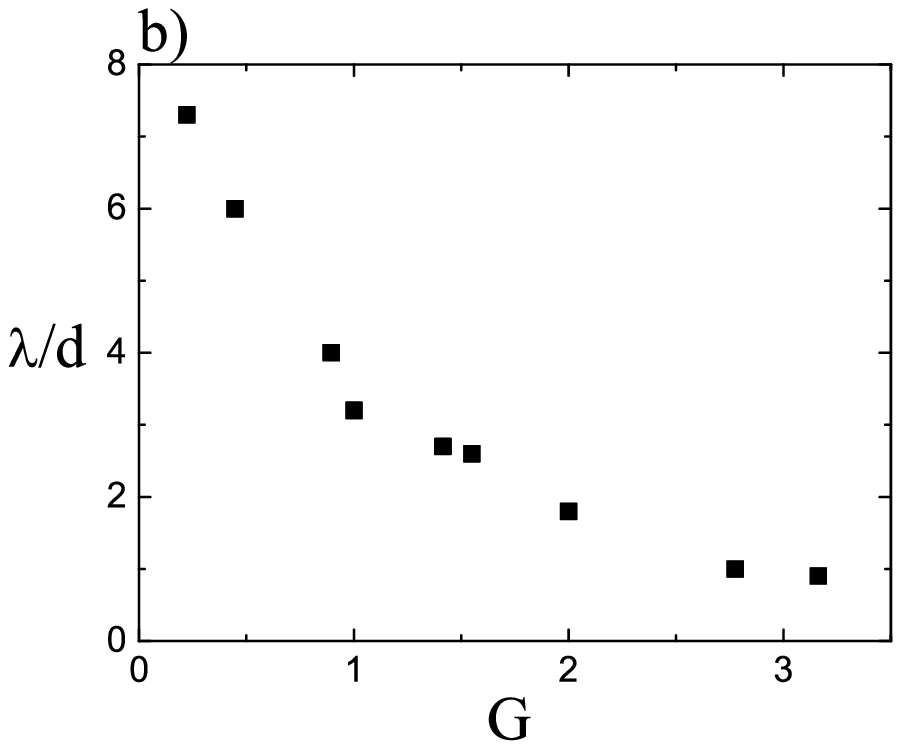}
            \caption{\label{Fig18}
            \textit{(Color online) Influence of $G$: (a) Velocity profiles
            (logarithmic scale)
            (b) Width $\lambda$ of the shear layer
            ($H/d = 40, e = 0.1, \mu =0.4, I_g=0.015
            $)}}
        \end{center}
    \end{figure}

The measured velocity profiles $v_x(y)$ (Fig.~\ref{Fig18} (a))
indicate a shear localization near the moving wall. Those profiles
are approximately exponential in the upper part (see
\cite{Volfson03}), $v_{x}(y) \simeq V \exp(-\frac{H-y}{\lambda})$,
from which we estimate the width of the shear layer $\lambda$.
Contrarily to the annular shear or vertical chute
geometries~\cite{Gdr04}, where this length is always of the order
of five to ten grain diameters, this length strongly varies as a
function of $G$ (Fig.~\ref{Fig18} (b)) and diverges when $G \to
0$, which means that the flow becomes homogeneous in the limit of
small gravity. The solid fraction increase at distance from the
moving wall and is very high in the quasi-static region.

\subsection{Dilatancy and friction laws}

\begin{figure}[!htb]
        \begin{center}
            \includegraphics*[width=8cm]{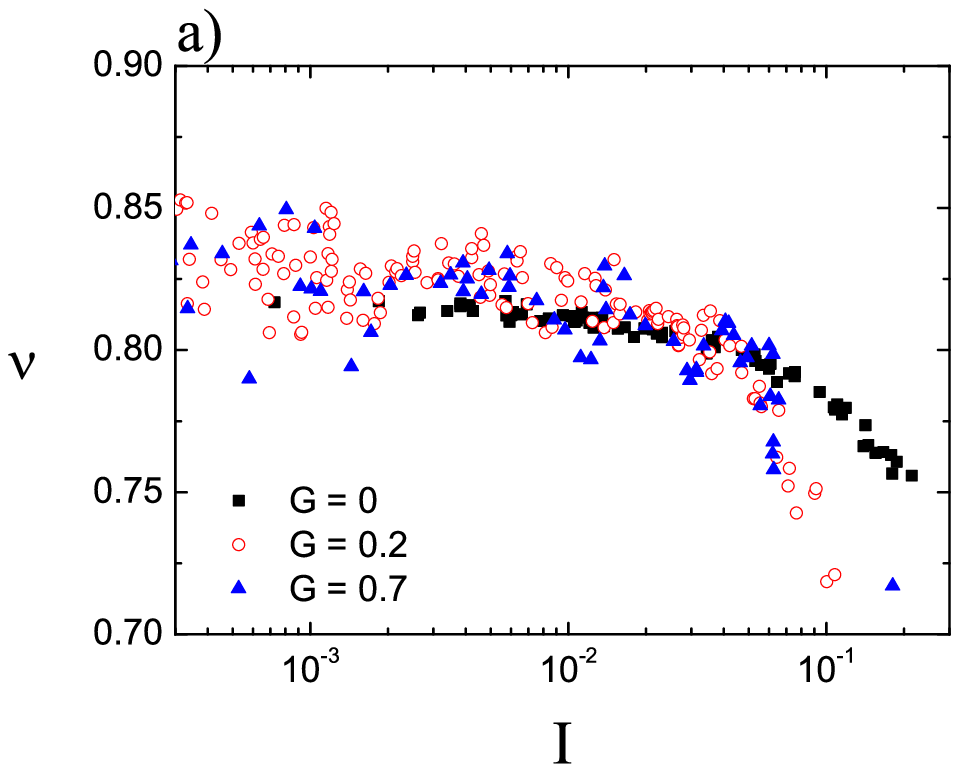}
            \includegraphics*[width=8cm]{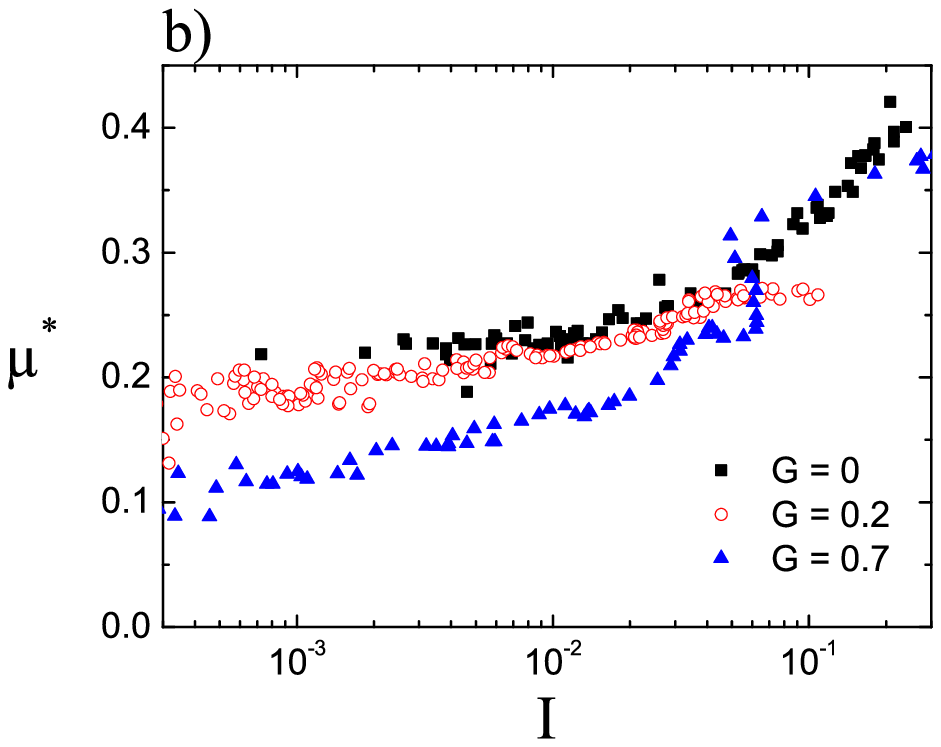}
            \caption{\label{Fig19}
            \textit{(Color online) Influence of $G$:  (a) Dilatancy law
            (b) Friction law (logarithmic scale)
            ($e = 0.1, \mu = 0.4, I_g = 0.015$).}}
        \end{center}
    \end{figure}

For each value of $G$, we deduce from $\nu(y)$, $\dot \gamma(y)$,
$P(y)$ and $S(y)$ the variations of the local solid fraction and
of the effective friction coefficient as a function of the local
inertial number (Fig.~\ref{Fig19} (a) et (b)). A single simulation
provides the whole curve, but the data are noisy since they result
from local measurements, which are not averaged over the whole
layer. The tendency is analogous to the no-gravity case: decrease
of $\nu$ from $\nu_{max}$, and increase of $\mu^*$ from
$\mu^*_{min}$ when $I$ increases. However, as $G$ increases,
$\nu_{max}$ increases while $\mu^*_{min}$ decreases. The $a$ and
$b$ factors are affected as well.

We have also measured the influence of $G$ on the contact network
and on the fluctuations~\cite{Dacruz04a} : $G$ has nearly no
influence on the $Z(I)$ and $M(I)$ dependencies; the translation
and rotation scaling laws~\eqref{eqn:fluct} remain valid but the
exponents $\alpha$ and $\beta$ increase respectively from $0.5$ to
$0.6$ and from $0.3$ to $0.5$.

    \subsection{Comments}

Those measurements indicate that the measurements performed in
homogeneous shear flows are not sufficient to describe other flow
geometries. Then, the inertial number is not the only relevant
quantity to describe the shear flow. We must take into account
other quantities, such as the gravity number in plane shear flows
with gravity. At this stage, we are not able to explain
quantitatively the observations which we have briefly quoted, such
as the nearly exponential shear localization, or the decrease of
$\mu^*_{min}$ with $G$. Following our explanation of the friction
law in the homogeneous case (see Sec.~\ref{sec:frictionlaw}), we
suggest that this could be related to the evolution of the
anisotropy of the contact network: as $G$ increases, the
distribution $\zeta_N(\phi)$ would increase for $0 \le \phi \le
\pi/2$.

\section{Conclusion}

We now summarize our conclusions. We have considered the simplest
flow geometry (plane shear without gravity), where the stress
distribution is homogeneous. Using molecular dynamics simulation,
we have submitted a dense assembly of frictional, inelastic disks,
to a given pressure and shear rate. We have observed steady
homogeneous shear flows, which become intermittent in the
quasi-static regime. We have shown that, in the limit of rigid
grains, the shear state is determined by a single dimensionless
number, called inertial number $I$, which describes the ratio of
inertial to pressure forces. Small values of $I$ correspond to the
quasi-static critical state regime of soil mechanics, while large
values of $I$ correspond to the dynamic collisional regime of the
kinetic theory. When $I$ increases in the intermediate regime, we
have measured an approximately linear decrease of the solid
fraction from the maximum packing value, and an approximately
linear increase of the effective friction coefficient from the
static internal friction value. From those dilatancy and friction
laws, we have deduced a visco-plastic constitutive law, with a
plastic Coulomb term and a viscous Bagnold term. We have also
measured scaling laws for the relative velocity fluctuations as a
function of $I$. We have shown that the mechanical characteristics
of the grains (restitution, friction and elasticity) have a very
small influence in the intermediate regime. We have shown that the
origin of the friction law relies in the fluctuating rotations of
the grains and on the angular distribution of contact forces.

The study of plane shear flows with gravity has shown that the
generalization to other heterogeneous flow geometries is not
straightforward. However, the study of annular shear flows
~\cite{Dacruz04a} and inclined plane flows~\cite{Dacruz04a,
Dacruz05, Lois05} reveal the same qualitative tendencies. Other
flow geometries such as vertical chute, rotating drum and heap
flow should be analyzed in the same way~\cite{Gdr04}.

We notice that a generalization of those ideas to steady uniform
shear flows of cohesive granular materials has been
successful~\cite{Rognon05a}.

In this paper, we have restricted our attention to velocity
controlled shear flows, so that it was not possible to study the
flow threshold. We think that it was evidenced indirectly through
the appearance of intermittencies for small enough
$I$~\cite{Mills05}. A specific study of the jamming mechanisms
should be performed by controlling the shear stress, either in
plane shear flows~\cite{Volfson03} or down inclined
planes~\cite{Dacruz04a, Dacruz05}.

\section*{Acknowledgments}

We gratefully acknowledge Lydéric Bocquet, Philippe Coussot, Ivan
Iordanoff, Jim Jenkins, Pierre Mills and Olivier Pouliquen for
many interesting discussions at various stages of this study.

Laboratoire des Matériaux et Structures du Génie Civil is a joint
laboratory, depending on Laboratoire Central des Ponts et
Chaussées, Ecole Nationale des Ponts et Chaussées and Centre
National de la Recherche Scientifique.

\bibliography{./dacruz}

\end{document}